\documentclass[11pt,a4paper]{article}
\usepackage[utf8]{inputenc}
\usepackage{authblk}

\RequirePackage[T1]{fontenc}

\RequirePackage{graphicx}
\RequirePackage{mathptmx}      
\RequirePackage{flushend}
\RequirePackage[numbers,sort&compress]{natbib}
\RequirePackage[colorlinks,citecolor=blue,urlcolor=blue,linkcolor=blue]{hyperref}

\RequirePackage[utf8]{inputenc}

\newcommand\sss{\mathchoice%
{\displaystyle}%
{\scriptstyle}%
{\scriptscriptstyle}%
{\scriptscriptstyle}%
}

\newcommand{\pT}{p_{\sss \rm T}}
\newcommand{\pt}{\pT}
\newcommand\as{\alpha_{\sss\rm S}}
\newcommand{\nloqcd}{\rm NLO_{\rm QCD}}
\newcommand{\nloew}{\rm NLO_{\rm EW}}
\newcommand{\qcdqedps}{\rm PS_{{\rm QCD},{\rm QED}}}
\newcommand{\qcdps}{\rm PS_{{\rm QCD}}}
\newcommand{\RES}{{\tt POWHEG-BOX-RES}}
\newcommand{\BOX}{{\tt POWHEG-BOX}}
\newcommand{\BOXX}{{\tt POWHEG-BOX-V2}}
\newcommand{\mllcut}{M_{\ell\ell}^{\rm\sss cut}}
\newcommand{\mllsupp}{{M_{\ell\ell}^{\rm\sss supp}}}
\newcommand{\htsupp}{{H_{T}^{\rm\sss supp}}}
\newcommand{\dontshow}[1]{}
\newcommand{\mathd}{\mathrm{d}}

\def\({\left(} 
\def\){\right)} 

\begin{document}\sloppy

\title{NLO QCD+NLO EW corrections to diboson production matched to
  parton shower}

\author[1]{Mauro Chiesa\thanks{\href{mailto:mauro.chiesa@lapth.cnrs.fr}{mauro.chiesa@lapth.cnrs.fr}}}
\author[2]{Carlo Oleari\thanks{\href{mailto:carlo.oleari@mib.infn.it}{carlo.oleari@mib.infn.it}}}
\author[1]{Emanuele Re\thanks{\href{mailto:emanuele.re@lapth.cnrs.fr}{emanuele.re@lapth.cnrs.fr}}}

\affil[1]{LAPTh, Universit\'e Grenoble Alpes, Universit\'e Savoie Mont Blanc, CNRS, 74940 Annecy, France}
\affil[2]{Universit\`a di Milano\,-\,Bicocca and INFN, Sezione di Milano\,-\,Bicocca, Piazza della Scienza 3, 20126 Milano, Italy}


\maketitle

\begin{abstract}
We present the matching of NLO QCD and NLO EW corrections to parton
showers for vector-boson pair production at the LHC. We consider
leptonic final states, including resonant and non-resonant diagrams,
spin correlations and off-shell effects. Our results are obtained
interfacing the {\sc Recola2-Collier} one-loop provider with the
\RES{} framework. We discuss our implementation, we validate it at
fixed order, and we show our final results matched to parton shower. A
by-product of our work is also a general interface between {\sc
  Recola2-Collier} and \RES{}.  This is the first time that EW and QCD
corrections to diboson production are consistently matched to parton
showers.
\end{abstract}

\section{Introduction}
\label{sec:intro}

The pair production of massive vector bosons at the LHC ($pp\to VV'$, with
$VV'=\{W^+ W^-$, $ZZ$, $W^\pm Z\}$) is among the most studied Standard
Model~(SM) processes, both as a signal on its own and as a background to
physics beyond the Standard Model~(BSM) and Higgs searches. Electroweak boson
pair production involving at least a $W$ boson in the final state ($W^+W^-$
and $ZW^\pm$) is important for collider phenomenology because it is sensitive
to the $ZWW$ gauge-boson self interaction, and therefore precision
measurements of $VV'$ processes provide a test of the electroweak gauge
structure. These precision tests are usually carried out by setting bounds on
the allowed size of anomalous trilinear gauge
couplings~(aTGCs)~\cite{Hagiwara:1986vm}, although several other ideas have
been proposed to study effects due to BSM physics with $VV$ final
states~\cite{Frye:2015rba, Butter:2016cvz, Zhang:2016zsp, Green:2016trm,
  Baglio:2017bfe, Falkowski:2016cxu, Panico:2017frx, Franceschini:2017xkh,
  Baglio:2018bkm, Baglio:2019uty, Liu:2018pkg, Azatov:2019xxn}.  Diboson
production is also a background for several searches, notably those involving
an heavy resonance decaying to a pair of gauge bosons. In particular, $pp\to
W^+W^-$ and $pp\to ZZ$ are irreducible background for Higgs production, when
the Higgs boson decays to gauge bosons.

For all the above reasons, it is essential to make accurate predictions for
vector boson pair production processes. Among the possible final states, the
one where four leptons are present is probably the most interesting one, as
it allows precise measurements, due to its clear signature. In the rest of
this work, including this section, we will only discuss final states with
four leptons, although we will often use the $VV'$ intermediate state as a
shorthand notation to identify the full process. We will give more details on
the exact leptonic final state, and approximations made, only where needed.

The status of predictions at fixed-order in the strong~($\as$) and/or
electroweak~($\alpha$) coupling for diboson production with leptonic decays
is rather advanced. As far as QCD corrections are concerned, the
next-to-leading-order~(NLO) QCD corrections for the process
$q\bar{q}^{(\prime)}\to VV'$ (with leptonic decays, interference and
off-shell effects fully taken into account) were obtained in
Refs.~\cite{Ohnemus:1994ff, Dixon:1999di, Campbell:1999ah,
  Campbell:2011bn}. The NNLO QCD corrections were computed more recently, in
Refs.~\cite{Grazzini:2015hta, Grazzini:2016swo, Grazzini:2016ctr,
  Grazzini:2017ckn, Heinrich:2017bvg, Kallweit:2018nyv}.\footnote{Most of
  these results have been obtained using the {\tt MATRIX}
  framework~\cite{Grazzini:2017mhc}.} The loop-induced processes $gg\to
W^+W^-$ and $gg\to ZZ$ contribute to the final state at the same order in
$\as$ as the NNLO corrections to the $q\bar{q}$ initial state. They have been
computed at LO in Refs.~\cite{Binoth:2005ua, Zecher:1994kb, Binoth:2006mf},
and the relative NLO corrections~($\mathcal{O}(\as^3)$) were obtained more
recently, in Refs.~\cite{Caola:2015rqy, Caola:2015psa, Grazzini:2018owa,
  Grazzini:2020stb}.

Although electroweak~(EW) corrections are usually small for total cross
sections, they can have an impact on differential distributions.  Typically
this is the case for large invariant-mass or transverse-momentum~($\pt$)
distributions, due to the so-called EW Sudakov logarithms. Sizeable effects
due to radiative photons are also visible, for leptonic observables, near
resonances or kinematic thresholds. NLO EW corrections to the
$q\bar{q}^{(\prime)}\to VV'$ processes (with leptonic decays and interference
effects) were computed in Ref.~\cite{Biedermann:2016guo, Biedermann:2016yvs,
  Biedermann:2016lvg, Biedermann:2017oae, Kallweit:2017khh}, improving on the
results obtained for stable vector bosons in~\cite{Bierweiler:2012kw,
  Bierweiler:2013dja, Baglio:2013toa}. For on-shell $W^+W^-$ production,
subleading EW Sudakov corrections at next-to-next-to-leading
logarithmic~(NNLL) accuracy were considered in Ref.~\cite{Kuhn:2011mh}.

In the context of all-order computations in QCD, the NNLL-accurate results
for the transverse-momentum distribution of the leptonic final state arising
from $pp\to VV'$ production were obtained in Ref.~\cite{Meade:2014fca,
  Grazzini:2015wpa, Becher:2019bnm}, whereas jet-veto logarithms were
resummed at NNLL in Ref.~\cite{Becher:2014aya, Dawson:2016ysj,
  Arpino:2019fmo}. All these resummed results were matched to the inclusive
NLO (or NNLO) total cross sections. The most accurate predictions were
obtained in Ref.~\cite{Kallweit:2020gva}: the transverse momentum of the
$W^+W^-$ system was resummed at NNLO+N3LL accuracy, the jet-veto logarithms
were resummed at NNLO+NNLL, and a joint resummation for the $W^+W^-$ $\pt$
spectrum in presence of a jet veto was performed at NNLO+NNLL.

The matching of fixed-order computations in QCD with parton showers~(PS)
algorithms~(NLO-QCD+PS) is well established with the {\sc
  MC@NLO}~\cite{Frixione:2002ik} and the {\sc POWHEG}~\cite{Nason:2004rx,
  Frixione:2007vw} matching algorithms, that are implemented in the public
software {\tt MadGraph5\_aMC@NLO}~\cite{Alwall:2014hca} and
\BOX{}~\cite{Alioli:2010xd}. Variants of these algorithms are also available
within the {\tt Sherpa} generator~\cite{Hoeche:2011fd, Bothmann:2019yzt} and
in {\tt Herwig7}~\cite{Bellm:2015jjp}, through the {\tt MatchBox}
framework~\cite{Platzer:2011bc}. Dedicated studies of $pp\to VV'$ production
at NLO-QCD+PS, with full leptonic decays and including resonant and
non-resonant diagrams as well as spin correlations and off-shell effects
exactly, were performed in Ref.~\cite{Melia:2011tj, Frederix:2011ss,
  Cascioli:2013gfa, Nason:2013ydw, Alioli:2016xab, Baglio:2018bkm,
  Baglio:2019uty}.  Notably, starting from the NLO+PS merging of $pp\to
W^+W^-$ and $pp\to W^+W^-+j$~\cite{Hamilton:2016bfu} obtained through the
{\sc MiNLO} approach~\cite{Hamilton:2012rf}, NNLO-QCD+PS results for the
$pp\to W^+W^-$ process were obtained in Ref.~\cite{Re:2018vac}.

As far as the combination of QCD and EW results at fixed-order is concerned,
this has been studied at NLO QCD + NLO EW accuracy in
Ref.~\cite{Kallweit:2017khh}, and more recently, at NNLO QCD + NLO EW in
Ref.~\cite{Kallweit:2019zez}.

In this work we combine the NLO QCD corrections and the NLO EW ones, and
match them to parton shower for the first time. Our underlying NLO
computation is performed combining the exact $\mathcal{O}(\as)$ and the exact
$\mathcal{O}(\alpha)$ effects in an additive way, and it is matched to a
complete PS algorithm where both QCD and QED emissions are simulated. We will
discuss how the matching is performed in Sect.~\ref{sec:calc}. Here we stress
that our results are the first ones where the matching is achieved exactly
for diboson production.  In previous publications, as, for instance, for some
of the results presented in Ref.~\cite{Kallweit:2017khh}, QED corrections
were included only via the PS, after having subtracted from the hard matrix
elements, and in an approximate manner, the QED effects due to radiative
photon emission, while keeping the Sudakov logarithms of pure weak origin,
arising from virtual $W$ and $Z$ boson exchange.\footnote{This is the
  ``$\mbox{EW}_{\rm VI}$'' approximation introduced in
  Ref.~\cite{Kallweit:2015dum}.}  More recently, the same approximation was
used by the authors of Ref.~\cite{Brauer:2020kfv}, where a NLO-QCD+PS merging
of the $pp\to W^+W^-$ and $pp\to W^+W^-+j$ processes was achieved keeping
weak corrections and approximate QED effects.

We also remind the reader that, at fixed-order, a description of mixed terms
can be obtained via a factorized ansatz, i.e.~multiplying differential NLO
QCD cross sections by EW correction factors, as done, for instance, in
Ref.~\cite{Kallweit:2017khh}.
In our approach, mixed $\mathcal{O}(\as^n \alpha^m)$ terms are generated
through the exponentiation of QED and QCD radiation in the {\sc POWHEG}
Sudakov form factor, and the mixed terms generated in this way are then only
correct in the QCD and QED collinear limits.

The paper is organized as follows: in Sect.~\ref{sec:calc} we describe the
details of our computation, in Sect.~\ref{sec:param} we list the parameters
and cuts used throughout this work, and in Sect.~\ref{sec:checks} we discuss
the validation of our implementation. In Sect.~\ref{sec:nlops} we show a
selection of the new results, i.e.~the matching of NLO EW + NLO QCD
corrections to parton shower. We summarize our work in
Sect.~\ref{sec:conclusions}.

In the rest of this manuscript we will use the shorthand $\nloqcd$ and
$\nloew$ to denote NLO accuracy in the $\as$ and $\alpha$ perturbative
expansion, respectively. We use the notation $\nloqcd$ + $\nloew$ to denote
the additive combination of the hard matrix elements (in the {\sc POWHEG}
$\bar{B}$ function).

\section{Details of the calculation}
\label{sec:calc}

In this paper we consider the processes
\begin{eqnarray}
\label{eq:processes}  
pp &\to& e^+\nu_e \mu^- \overline{\nu}_\mu\,, \nonumber\\
pp &\to& \mu^+\nu_\mu e^- e^+\,, \nonumber\\
pp &\to& \mu^+\mu^- e^- e^+\,.   
\end{eqnarray}
We stress that the full matrix elements for four fermion production are used
and no on-shell or double pole approximation is employed. In the following
the three processes will be dubbed as $WW$, $WZ$, and $ZZ$ production, and,
collectively, as ``diboson production''. Although we will show results only
for $W^+Z$ production, our code is fully general and $W^-Z$ production can be
generated as well.

\begin{figure*}
  \begin{center}
    \includegraphics[scale=0.7]{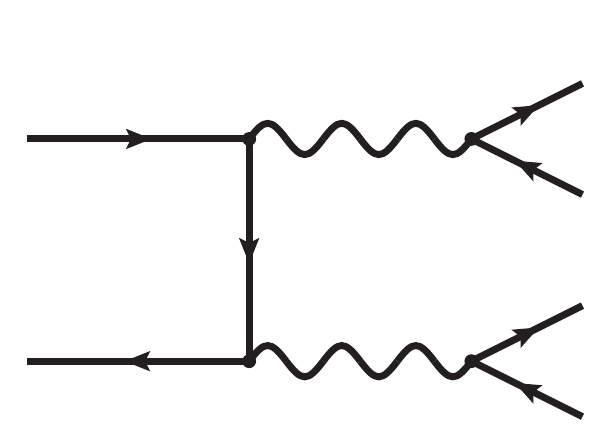}$\qquad\qquad$
    \includegraphics[scale=0.7]{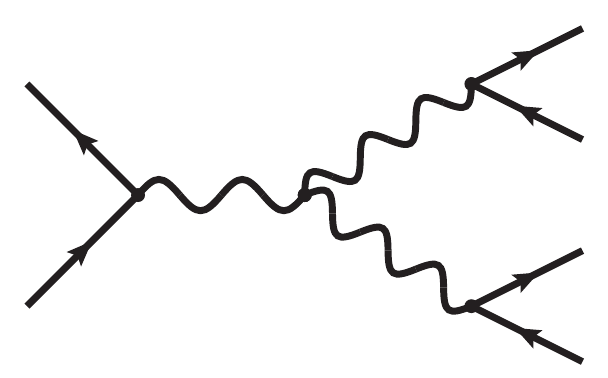}$\qquad\qquad$
    \includegraphics[scale=0.7]{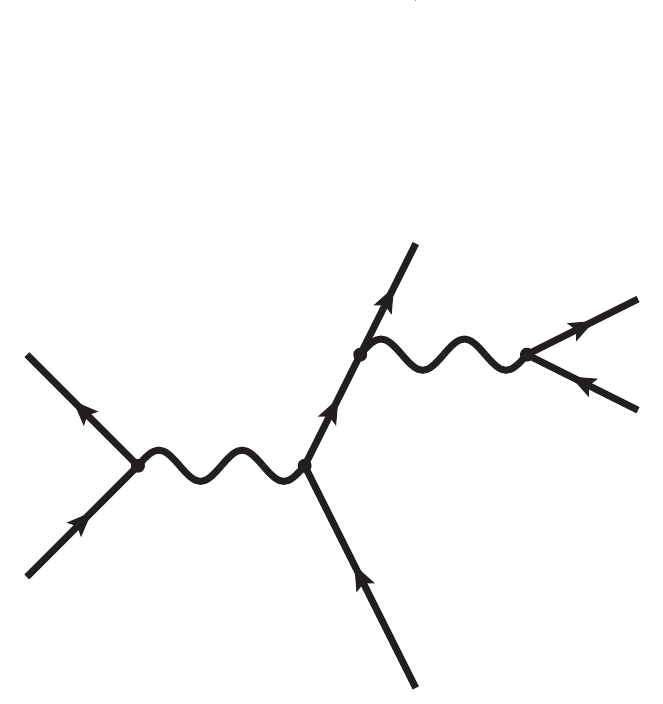}
    \caption{\label{feyndiags} Representative Feynman diagrams for the
      possible classes of resonance histories contributing at LO.}
  \end{center}
\end{figure*}
 
The calculation of the $\nloqcd$ + $\nloew$ corrections to diboson production
matched to QCD and QED parton shower presented in this paper is performed in
the \RES{} framework~\cite{Jezo:2015aia}, which is a framework designed to
simulate processes involving intermediate decaying resonances with NLO+PS
accuracy. It is a new implementation of the {\sc POWHEG}
method~\cite{Nason:2004rx, Frixione:2007vw} that overcomes the limitations of
the \BOX{} framework~\cite{Alioli:2010xd}.  It has been used in
Ref.~\cite{Jezo:2016ujg} to simulate the process $pp\to b\bar{b} \ell
\bar{\ell}\nu\bar{\nu}$ with $\nloqcd$+PS accuracy, thereby achieving, for
the first time, a fully-consistent treatment of $t\bar{t}$ and $Wt$
production with two leptonic decays, in Ref.~\cite{Granata:2017iod} to
compute the processes $pp\to HV$ and $pp\to HVj$ production~($V=W,Z$) with
$\nloqcd$ + $\nloew$+PS accuracy, and in Ref.~\cite{Chiesa:2019ulk} to
compute the $\nloew$+PS corrections to $pp\to \ell\ell'\nu\nu'jj$.  In
Ref.~\cite{CarloniCalame:2016ouw}, a simplified version of the \RES{}
algorithm has been implemented also in the {\sc W\_ew-BMNNP} and {\sc
  Z\_ew-BMNNPV} packages~\cite{Barze:2012tt, Barze:2013fru,
  CarloniCalame:2016ouw, Chiesa:2019nqb} of \BOXX{}, in order to simulate
neutral and charged Drell-Yan production with $\nloqcd$ + $\nloew$+PS
accuracy in a fully-consistent manner.\footnote{A $\nloqcd$ + $\nloew$+PS
  implementation of the charged Drell-Yan case obtained with the \BOXX{}
  algorithm was obtained in Ref.~\cite{Bernaciak:2012hj}.} A fully
independent calculation of $\nloqcd$ + $\nloew$+PS corrections to Drell-Yan
production was performed also in Ref.~\cite{Muck:2016pko}.

The basic {\sc POWHEG} cross section formula is given by
\begin{equation}
  \label{eq:dsigma}
    d\sigma = \bar{B}(\Phi_B) \, \mathd\Phi_B \left[ \Delta_{\pT^0}\!\(\Phi_B\)\right.
      \left.{}
      + \frac{ R_{\rm \sss QCD }(\Phi_B,
        \Phi_{\rm \sss rad})
        + R_{\rm \sss EW }(\Phi_B, \Phi_{\rm \sss rad}) }{ B (\Phi_B) }
      \Delta_{\pT}\!\(\Phi_B\)  \mathd \Phi_{\rm \sss rad} 
      \right]
\end{equation}
where
\begin{equation}
  \label{eq:Bbar} 
  \bar{B}(\Phi_B) = B (\Phi_B) + \left[V_{\rm\sss QCD}(\Phi_B) + V_{\rm\sss
      EW}(\Phi_B)\right]
   {} + \int \mathd \Phi_{\rm \sss rad} \left[ R_{\rm \sss QCD }(\Phi_B,
    \Phi_{\rm \sss rad})
    + R_{\rm \sss EW }(\Phi_B, \Phi_{\rm \sss rad}) \right]
\end{equation}
and $\Delta_{\pT}\!\(\Phi_B\)$ is the Sudakov form factor
\begin{equation}
  \label{eq:Sudakov} 
  \Delta_{\pT}\!\(\Phi_B\) = \Delta^{\rm \sss QCD }_{\pT}\!\(\Phi_B\)   \times
  \Delta^{\rm \sss EW }_{\pT}\!\(\Phi_B\) ,
\end{equation}
with   
\begin{eqnarray}
   \Delta^{\rm \sss QCD }_{\pT}\!\(\Phi_B\) &=&
  \exp \left[ - \!\!\int_{\pT}\!\!  \mathd \Phi_{\rm \sss rad} \frac{
        R_{\rm \sss QCD } (\Phi_B, \Phi_{\rm \sss rad})}{B(\Phi_B)}
    \right]\nonumber
  \\
\Delta^{\rm \sss EW }_{\pT}\!\(\Phi_B\) &=&
  \exp \left[ - \!\!\int_{\pT}\!\!  \mathd \Phi_{\rm \sss rad} \frac{
        R_{\rm \sss EW } (\Phi_B, \Phi_{\rm \sss rad})}{B(\Phi_B)}      
    \right]\nonumber
\end{eqnarray}
where $B$, $V$ and $R$ are the Born, the virtual and the real matrix
elements, and ``QCD'' and ``EW'' refer to the strong (order $\as$) and
electroweak (order $\alpha$) corrections. $\Phi_B$ and $\Phi_{\rm \sss rad}$
are the phase-space volumes for the Born and the radiation kinematics. The
integration region in the Sudakov form factor $\Delta_{\pT}$ covers the
phase-space volume where the transverse momentum of the radiated particle is
larger than $\pT$.  The Sudakov form factor is split into two terms,
constructed using the real contribution coming from QCD and photon radiation.
Following the {\sc POWHEG} method for generation of radiation, one generates
QCD or QED radiation from each singular region, and, at the end, the hardest
radiation is kept.
In case of initial-state radiation~(ISR), this implies a competition between
QED and QCD radiation from the initial-state quarks. The radiation from the
resonances is only of QED type, involving only photon emission off leptons.

The improvement of the algorithm implemented in \RES{} with respect to
\BOXX{} is twofold.  We summarize it briefly in this paragraph, and we refer
to Ref.~\cite{Jezo:2015aia} for more details. On the one hand, the
calculation of the NLO predictions needed for the event generation uses a
modified version of the FKS~\cite{Frixione:1995ms} algorithm for the
subtraction of the infrared~(IR) singularities, that takes into account the
resonant structure of the process under consideration through the concept of
``resonance history''. Not only does this modification improve the
integration stability, but it also allows one to generate the {\sc POWHEG}
hardest radiation preserving the intermediate resonance(s) virtuality
everywhere in the {\sc POWHEG} Sudakov, preventing shape distortions in the
matching to PS.  On the other hand, \RES{} can generate up to one ``hard''
radiation from each resonance (including the hard production process among
the resonances):\footnote{This corresponds to the so called {\tt allrad}
  scheme, first introduced in Ref.~\cite{Campbell:2014kua}.} the hardness of
each radiation is to be used as a veto scale for PS evolution of the
particles belonging to the resonance that emitted the considered radiation.
As discussed in Ref.~\cite{CarloniCalame:2016ouw}, the latter point is
crucial when computing the $\nloqcd$ + $\nloew$+PS corrections to observables
that are very sensitive to final-state QED radiation~(FSR QED) but rather
insensitive to initial-state QCD corrections~(ISR QCD).  This is the case,
for instance, of the dilepton invariant-mass distributions in the region
close to the nominal vector-boson masses, where the predictions, obtained
with and without the {\tt allrad} option, can differ at the percent level.

In order to implement the $\nloqcd$ and the $\nloew$ corrections to diboson
production in \RES{}, we had to define the list of all the contributing LO
and real partonic subprocesses together with the corresponding resonance
histories, and to provide the required Born, virtual, and real matrix
elements. We decided to code all the three classes of diboson-production
processes (namely, four charged leptons, three charged leptons plus one
neutrino, and two charged leptons plus two neutrinos) in the same {\sc
  POWHEG} package and let the user select the desired one from the input
card.

Concerning the resonance histories, we consider the $t$-channel ones with two
decaying vector bosons (Fig.~\ref{feyndiags}, left), the $s$-channel ones
involving triple gauge-boson interactions (for $WW$ and $WZ$ production,
Fig.~\ref{feyndiags}, center), and the peripheral ones involving the
$s$-channel production of a vector boson that decays into a dilepton pair
which radiates a second vector boson (Fig.~\ref{feyndiags}, right).  While
the first two classes of resonance histories are by far the dominant ones in
the typical event selections for diboson production, the third one can be
important if a more inclusive analysis is considered (this could be the case,
for instance, if the code is used to simulate background contributions to
other processes with four final-state leptons).

We found that the inclusion of all the resonance histories in
Fig.~\ref{feyndiags}, and in particular the peripheral ones, improves the
numerical stability of the calculation and strongly reduces the size of the
``remnant'' cross section.\footnote{We stress the fact that we always employ
  full matrix elements: the definition of the peripheral resonance histories
  only affects the way \RES{} performs the subtraction of the IR
  singularities and the integration. The concept of the ``remnant'' cross
  section in the \BOX{} codes was introduced in Ref.~\cite{Alioli:2010xd}.}
Moreover, since the PS preserves the virtualities of the unstable particles
written in the Les Houches~(LH) events, defining all the resonance histories
prevents a possible mismodeling of the treatment of events with a nested
resonance structure.

All the needed Born, virtual, and real matrix elements are computed using the
{\sc Recola2-Collier} one-loop provider. {\sc Recola2}~\cite{Actis:2012qn,
  Actis:2016mpe, Denner:2017vms, Denner:2017wsf} is a library for the fully
automated calculation of tree-level and one-loop matrix elements which relies
on the {\sc Collier}~\cite{Denner:2010tr, Denner:2002ii, Denner:2005nn,
  Denner:2016kdg} library for the reduction of the tensor integrals and the
evaluation of the scalar integrals coming from the one-loop diagrams. We use
the {\sc SM-2.2.2} Recola model file to compute the $\nloqcd$ + $\nloew$
corrections in the SM, but in principle our code can be easily generalized to
use any BSM Recola model file as far as the considered extension of the SM
does not involve any modification of the interactions between photons and
fermions or among partons (as such modifications might have an impact on the
IR subtraction performed by \BOX{} and on the event generation). We developed
a completely general interface between \RES{} and {\sc Recola2} that can be
used for other processes of interest.\footnote{While the interface to {\sc
    Recola2} is general, the current treatment of the $\nloew$ corrections in
  \RES{} is not, as it implies that each virtual process is in one-to-one
  correspondence with a LO process (so that it can be considered either as a
  NLO QCD correction or a NLO EW correction to the corresponding LO process),
  which in general is not the case for complicated processes.  See for
  instance the $\mathcal{O}(\alpha_S \alpha^6)$ and $\mathcal{O}(\alpha_S^2
  \alpha^5)$ corrections to $pp\to
  \ell\ell'\nu\nu'jj$~\cite{Biedermann:2017bss}.}

In order to deal with the presence of unstable particles, {\sc Recola2}
implements the complex-mass scheme~(CMS)~\cite{Denner:1999gp, Denner:2005fg,
  Denner:2006ic}. In this scheme, the $W$ and $Z$ boson masses are promoted
to complex numbers with the replacement $M_V^2 \to \mu_V^2=M_V^2-i\,\Gamma_V
M_V$, and all the parameters derived from the gauge-boson masses (like, for
instance, the sine of the weak mixing angle) get an imaginary part.

Concerning the calculation of EW corrections, {\sc Recola2} allows to
perform the renormalization of the UV singularities in the SM using
three possible input parameter schemes: $(G_\mu,M_W,M_Z)$,
$(\alpha(M_Z),M_W,M_Z)$, and $(\alpha_0,M_W,M_Z)$. The results shown
in the following are computed in the $(G_\mu,M_W,M_Z)$ scheme, but our
code gives the user the possibility to select any of the
renormalization schemes mentioned above.

In the calculation of diboson-production cross sections and/or distributions,
there are tree-level singularities coming from the presence of $s$-channel
photon propagators that can go on-shell. In order to prevent these
singularities, we impose both generation cuts and suitable phase-space
suppression factors~\cite{Alioli:2010qp}. For example, if we consider the
process $pp \to e^+e^-\mu^+\mu^-$, the generation cuts read:
\begin{equation}
  M_{e^+e^-} > \mllcut,\qquad M_{\mu^+\mu^-} > \mllcut,
  \label{eq:gencutzz}
\end{equation}
while the suppression factor is:
\begin{equation}
  \frac{M_{e^+e^-}^4}{\left[M_{e^+e^-}^2+\left(\mllsupp\right)^2\right]^2}
  \frac{M_{\mu^+\mu^-}^4}{\left[M_{\mu^+\mu^-}^2+\left(\mllsupp\right)^2\right]^2}, 
  \label{eq:gensuppzz}
\end{equation}
where $M_{e^+e^-}$ and $M_{\mu^+\mu^-}$ are the invariant masses of the
underlying-Born electronic and muonic pair, and the actual values of
$\mllcut$ and $\mllsupp$ should be chosen by the user and depend on the cuts
applied during the analysis. On top of the suppression factor in
Eq.~(\ref{eq:gensuppzz}), we also provide a suppression factor of the form:
\begin{equation}
  \frac{\left(\htsupp\right)^{-4}}{\left[\left(\htsupp\right)^{-2}+H_T^{-2} \right]^2}, 
  \label{eq:gensupphtzz}
\end{equation}
where $H_T$ is the scalar sum of the transverse momenta of the charged
leptons, in order to improve the generation efficiency in the typical
diboson-event selection. Also for this suppression factor, the actual value
of $\htsupp$ is chosen by the user.~\footnote{In the code, $\mllcut$,
  $\mllsupp$ and $\htsupp$ are represented by the variables {\tt mllcut},
  {\tt mllsupp} and {\tt htsupp}, respectively.} 
The suppression factors only affect the integrand function
and do not affect the physical cross sections and distributions
since they are counterbalanced by suitable event weights.

The $WZ$ production process features approximated radiation zeros, that could
result in a highly inefficient generation of radiation according to the
Sudakov form factor in Eq.~(\ref{eq:Sudakov}), when $B$ becomes very
small. For this reason, we activate the {\tt bornzerodamp}
option~\cite{Alioli:2010xd,Melia:2011tj} of the \RES{}, for all the three
processes at hand. When {\tt bornzerodamp} is on, the regions of phase space
where the real matrix elements are very far from their singular limit are
removed from the $\bar{B}$ function, and treated separately as
``remnants''. 

As a final remark, the contribution of the loop induced $gg \to ZZ$ and $gg
\to W^+W^-$ processes is not included in our calculation. Even though these
are $\mathcal{O}(\as^2)$ effects, their impact is not negligible because of
the size of the gluon PDF.  These processes can be computed, independently,
at LO+PS using tools like {\sc gg2ZZ} and {\sc gg2WW}~\cite{Binoth:2005ua,
  Binoth:2006mf}. $\nloqcd$+PS results were presented in
Ref.~\cite{Alioli:2016xab}.  Photon-induced processes are not included in our
calculation.  As illustrated in Refs.~\cite{Biedermann:2016guo,
  Biedermann:2016yvs, Biedermann:2016lvg, Biedermann:2017oae,
  Kallweit:2017khh, Kallweit:2019zez}, these contributions can be
phenomenologically relevant.
Dealing with initial-state photons requires extra features in the \RES{}
code, not available while we write. We plan to to include them in a future
release of our code.

\section{Input parameters and cuts}
\label{sec:param}

The input parameters used in the numerical simulations at $\sqrt{s}=13$~TeV
are the following:
\begin{equation}
  \begin{array}{rclrcl}
  M_H &=& 125 \;{\rm GeV}, & \Gamma_H &=& 4.097 \;{\rm MeV},     \\[0.8mm]
  M_{\rm top} &=& 173.2 \;{\rm GeV}, & \Gamma_{\rm top} &=& 1.369 \;{\rm GeV},  \\[0.8mm]
  M_W^{\rm \sss OS} &=& 80.385 \;{\rm GeV}, & \Gamma_W^{\rm \sss OS} &=& 2.085 \;{\rm GeV}, \\[0.8mm]
  M_Z^{\rm \sss OS} &=& 91.1876 \;{\rm GeV}, & \Gamma_Z^{\rm \sss OS} &=& 2.4952 \;{\rm GeV},  \\[0.8mm]
  G_\mu &=& 1.1663787 \times 10^{-5}\; {\rm GeV}^{-2}. &&       
  \end{array}
  \label{eq:params}
\end{equation}
All fermions are considered as massless, with the exception of the top
quark. For this reason, we only provide results for dressed leptons. The
on-shell values of the $W$ and $Z$ masses and widths are converted internally
to the corresponding pole values with the relations:
\begin{equation}
  M_V=\frac{M_V^{\rm \sss OS}}{\sqrt{1+\Big(\frac{\Gamma_V^{\rm \sss
          OS}}{M_V^{\rm \sss OS}}\Big)^2}}, \qquad 
  \Gamma_V=\frac{\Gamma_V^{\rm \sss OS}}{\sqrt{1+\Big(\frac{\Gamma_V^{\rm
          \sss OS}}{M_V^{\rm \sss OS}}\Big)^2}}. 
  \label{eq:ostopole}
\end{equation}
For $WZ$ and $ZZ$ production, we set the generation cut $\mllcut$ of
Eq.~(\ref{eq:gencutzz}) at 15~GeV, for each same-flavour opposite-charged
lepton pair, and we apply the suppression factors in
Eqs.~(\ref{eq:gensuppzz}) and~(\ref{eq:gensupphtzz}) with $\mllsupp = 30$~GeV
and $\htsupp = 4$~GeV. We checked that our results do not depend on these
technical parameters in the event selection under consideration.

The UV renormalization for the EW corrections is performed in the on-shell
scheme with input parameters $(G_\mu,M_W,M_Z)$ supplemented with the CMS for
the treatment of the unstable particles. The $\overline{ \rm MS}$ scheme is
used for the renormalization of the NLO QCD corrections. In the following,
the factorization and renormalization scales are set to $\mu=(M_V^{\rm \sss
  OS}+M_{V'}^{\rm \sss OS})/2$ (where $V,V'=W,Z$ are the vector bosons that
define the signature under consideration), for constant scales, or to the
invariant mass of the four-lepton system at the underlying-Born level, when
using running scales.  The results at NLO+PS accuracy are only computed for
running scales.

The Cabibbo-Kobayashi-Maskawa~(CKM) matrix is set to the identity
matrix. However, in the code, the user can select a non-trivial quark mixing
matrix: in this case, the NLO EW corrections are still computed with $V_{\rm
  CKM}=1$ and then multiplied by the actual CKM values coming from the LO
part of the amplitude as in Refs.~\cite{Barze:2012tt, Biedermann:2017oae}.

In order to make contact with the results of Ref.~\cite{Chiesa:2018lcs}, the
{\tt NNPDF23\_nlo\_as\_0118\_qed} PDF set~\cite{Ball:2012cx, Ball:2013hta,
  Ball:2014uwa} is used. However, the user can select any modern PDF
set~\cite{Ball:2017nwa, Harland-Lang:2014zoa, Hou:2019efy}.  The PDF
evolution as well as the evolution of the strong coupling constant is
provided by the {\sc LHAPDF6} library~\cite{Buckley:2014ana}.

As in Ref.~\cite{Chiesa:2018lcs}, we always use the same value of $\alpha$
(namely, the one derived from $G_\mu$, i.e.~$\alpha^{-1}\simeq 132.357$) both
for the LO couplings and for the coupling when computing the NLO corrections,
both real and virtual. This introduces a small mismatch when {\sc POWHEG} is
interfaced to the QED PS, since the PS uses $\alpha_0$ for the photon-fermion
coupling ($\alpha_0^{-1}=137.03599911$). On the one hand, this mismatch is
really small and hardly visible on the scale of our plots and, on the other
hand, we allow the user to define two different values of $\alpha$: one to be
used for the LO couplings and a second one (corresponding to $\alpha_0$) to
be used in the additional power of $\alpha$ in the EW corrections.  When this
option is selected, {\sc POWHEG} performs the subtraction of the IR
singularities using $\alpha_0$, while the virtual and real matrix element are
computed by {\sc Recola2} with a different value of $\alpha$ and then
rescaled by a factor $\alpha_0/\alpha$.

For all diboson-production processes, the $b$ quark is treated as massless,
both in the initial and final state, when present.  For $WW$ production, we
do not include the contribution of initial-state $b$ quarks, in order to
remove the real QCD channel $gb \to W^+W^-b$ which is enhanced by the
presence of the top-quark resonance, but is usually subtracted in
experimental analysis (single-top background).

We provide a dedicated interface for a consistent matching with the {\sc
  PYTHIA8.2}~\cite{Sjostrand:2006za, Sjostrand:2014zea} PS, that will
generate secondary QED and QCD emissions and finally convert partons into
hadrons. As we will explain in Sect.~\ref{sec:nlops}, a dedicated interface
is necessary because we use the {\tt allrad} scheme in {\sc POWHEG}.

In this paper we do not consider distributions involving jets, however, we
provide a template analysis that can use {\sc Fastjet}~\cite{Cacciari:2005hq,
  Cacciari:2011ma} to reconstruct them.

In order to make the discussion of the results easier, we use the same basic
event selection for all diboson-production processes:
\begin{equation}
  \pT^{\ell} > 10 \;{\rm GeV}, \qquad |y^\ell|<2.5, \qquad \Delta R(\ell,\ell') > 0.3,
  \label{eq:cuts}
\end{equation}
where $\ell$ and $\ell'$ are charged leptons, and $\Delta R$ is the
separation in rapidity and azimuthal angle. For $pp \to e^+e^-\mu^+\mu^-$ and
$pp \to e^+e^-\mu^+\nu_{\mu}$ we also impose a leptonic mass window around
the $Z$-boson mass:
\begin{equation}
  80 \; {\rm GeV} < M(\ell^+\ell^-) < 110 \; {\rm GeV}, \qquad \ell=e,\, \mu.
  \label{eq:cuts2}
\end{equation}
Both muons and electrons are dressed: photons are recombined with charged
leptons if their angular distance $\Delta R(\ell,\gamma)$ is less than 0.1.

\section{Cross-checks and validation}
\label{sec:checks}

\begin{table}
  \begin{center}
  \begin{tabular}{ | c | c | c | c |}
    \cline{2-4}
 \multicolumn{1}{c}{}    &\multicolumn{3}{|c|}{$\sigma$ LO [pb]\qquad  fixed
   scales\phantom{\Large|}    } \\
    \cline{2-4}
 \multicolumn{1}{c|}{\phantom{\Large|}}    & $WW$  &  $ZZ$  &  $WZ$ \\
     \cline{1-4}
    MC\phantom{\Large|}   & 0.48003(3) & 0.0099340(4) &  0.026678(2)   \\
     \cline{1-4} 
    POWHEG\phantom{\Large|}   & 0.48003(3) & 0.0099344(3) &  0.02668(2)   \\
    \cline{1-4}  
  \end{tabular}
  \end{center}  
  \caption{\label{tab:LOfix} Integrated cross sections for the processes $pp
    \to e^+\nu_e \mu^- \overline{\nu}_\mu$, $pp \to \mu^+\mu^- e^- e^+$, and
    $pp \to \mu^+\nu_\mu e^- e^+$ at LO under the event selection in
    Eqs.~(\ref{eq:cuts}) and~(\ref{eq:cuts2}).  The factorization scale is
    set to $\mu=(M_V^{\rm \sss OS}+M_{V'}^{\rm \sss OS})/2$, ($V,V'=W,Z$).}
\end{table}

\begin{table}
  \begin{center}
    \begin{tabular}{ | c | c | c | c |}
    \cline{2-4}
 \multicolumn{1}{c}{}    &\multicolumn{3}{|c|}{$\sigma$ LO [pb]\qquad
   running scales\phantom{\Large|}    } \\
    \cline{2-4}
\multicolumn{1}{c|}{\phantom{\Large|}}    & $WW$  &  $ZZ$  &  $WZ$ \\
     \cline{1-4}
    MC \phantom{\Large|}  & 0.51961(3)  & 0.0107362(4) &  0.028547(2)   \\
    \cline{1-4} 
    POWHEG \phantom{\Large|}  & 0.51963(3)   & 0.0107367(3) &  0.02854(2)   \\
    \cline{1-4} 
    \end{tabular}
    \end{center}  
  \caption{\label{tab:LOrun} Integrated cross sections for the processes $pp
    \to e^+\nu_e \mu^- \overline{\nu}_\mu$, $pp \to \mu^+\mu^- e^- e^+$, and
    $pp \to \mu^+\nu_\mu e^- e^+$ at LO under the event selection in
    Eqs.~(\ref{eq:cuts}) and~(\ref{eq:cuts2}).  The factorization scale is
    set to the invariant mass of the four-fermion system.}
  \end{table}

\begin{table}
  \begin{center}
    \begin{tabular}{ | c | c | c | c |}
    \cline{2-4}
 \multicolumn{1}{c}{}&\multicolumn{3}{|c|}{$\sigma$ $\nloqcd$ [pb]\qquad  fixed scales\phantom{\Large|}    } \\
    \cline{2-4}
\multicolumn{1}{c|}{\phantom{\Large|}}    & $WW$  &  $ZZ$  &  $WZ$ \\
     \cline{1-4}
    MC \phantom{\Large|}  & 0.7221(1)  & 0.013421(2) &  0.048361(8)   \\
    \cline{1-4} 
    POWHEG \phantom{\Large|}  & 0.7224(2)  & 0.013420(2) &  0.04835(9)    \\
    \cline{1-4} 
    \end{tabular}
  \end{center}
  \caption{\label{tab:NLOQCDfix} Integrated cross sections for the processes
    $pp \to e^+\nu_e \mu^- \overline{\nu}_\mu$, $pp \to \mu^+\mu^- e^- e^+$,
    and $pp \to \mu^+\nu_\mu e^- e^+$ at NLO QCD under the event selection in
    Eqs.~(\ref{eq:cuts}) and~(\ref{eq:cuts2}).  The factorization and
    renormalization scales are set to $\mu=(M_V^{\rm \sss OS}+M_{V'}^{\rm \sss OS})/2$,
    ($V,V'=W,Z$).}
\end{table}

\begin{table}
  \begin{center}
    \begin{tabular}{ | c | c | c | c |}
    \cline{2-4}
 \multicolumn{1}{c}{}&\multicolumn{3}{|c|}{$\sigma$ $\nloqcd$ [pb]\qquad  running scales\phantom{\Large|}    } \\
    \cline{2-4}
\multicolumn{1}{c|}{\phantom{\Large|}}    & $WW$  &  $ZZ$  &  $WZ$ \\
     \cline{1-4}
    MC \phantom{\Large|}    & 0.7012(1) & 0.013236(2) &  0.045585(7) \\
    \cline{1-4} 
    POWHEG \phantom{\Large|}   & 0.70133(8) & 0.013234(2) &  0.045596(9) \\
    \cline{1-4} 
    \end{tabular}
  \end{center}
  \caption{\label{tab:NLOQCDrun} Integrated cross sections for the processes
    $pp \to e^+\nu_e \mu^- \overline{\nu}_\mu$, $pp \to \mu^+\mu^- e^- e^+$,
    and $pp \to \mu^+\nu_\mu e^- e^+$ at NLO QCD under the event selection in
    Eqs.~(\ref{eq:cuts}) and~(\ref{eq:cuts2}).  The factorization and
    renormalization scales are set to the invariant mass of the four-fermion
    system.}
\end{table}

\begin{table}
  \begin{center}
    \begin{tabular}{ | c | c | c | c |}
    \cline{2-4}
 \multicolumn{1}{c}{}&\multicolumn{3}{|c|}{$\sigma$ $\nloew$ [pb]\qquad  fixed scales\phantom{\Large|}    } \\
    \cline{2-4}
\multicolumn{1}{c|}{\phantom{\Large|}}    & $WW$  &  $ZZ$  &  $WZ$ \\
     \cline{1-4}
    MC \phantom{\Large|}    & 0.46961(9)  & 0.0088732(8) &  0.025281(8)  \\
    \cline{1-4} 
    POWHEG \phantom{\Large|}   & 0.46953(4)   & 0.008874(1)  &  0.025279(5) \\
    \cline{1-4} 
    \end{tabular}
  \end{center}
  \caption{\label{tab:NLOEW} Integrated cross sections for the processes $pp
    \to e^+\nu_e \mu^- \overline{\nu}_\mu$, $pp \to \mu^+\mu^- e^- e^+$, and
    $pp \to \mu^+\nu_\mu e^- e^+$ at NLO EW under the event selection in
    Eqs.~(\ref{eq:cuts}) and~(\ref{eq:cuts2}).  The factorization scale is
    set to $\mu=(M_V^{\rm \sss OS}+M_{V'}^{\rm \sss OS})/2$, ($V,V'=W,Z$).}
\end{table}

\begin{figure*}
  \begin{center}  
    \includegraphics[scale=0.8]{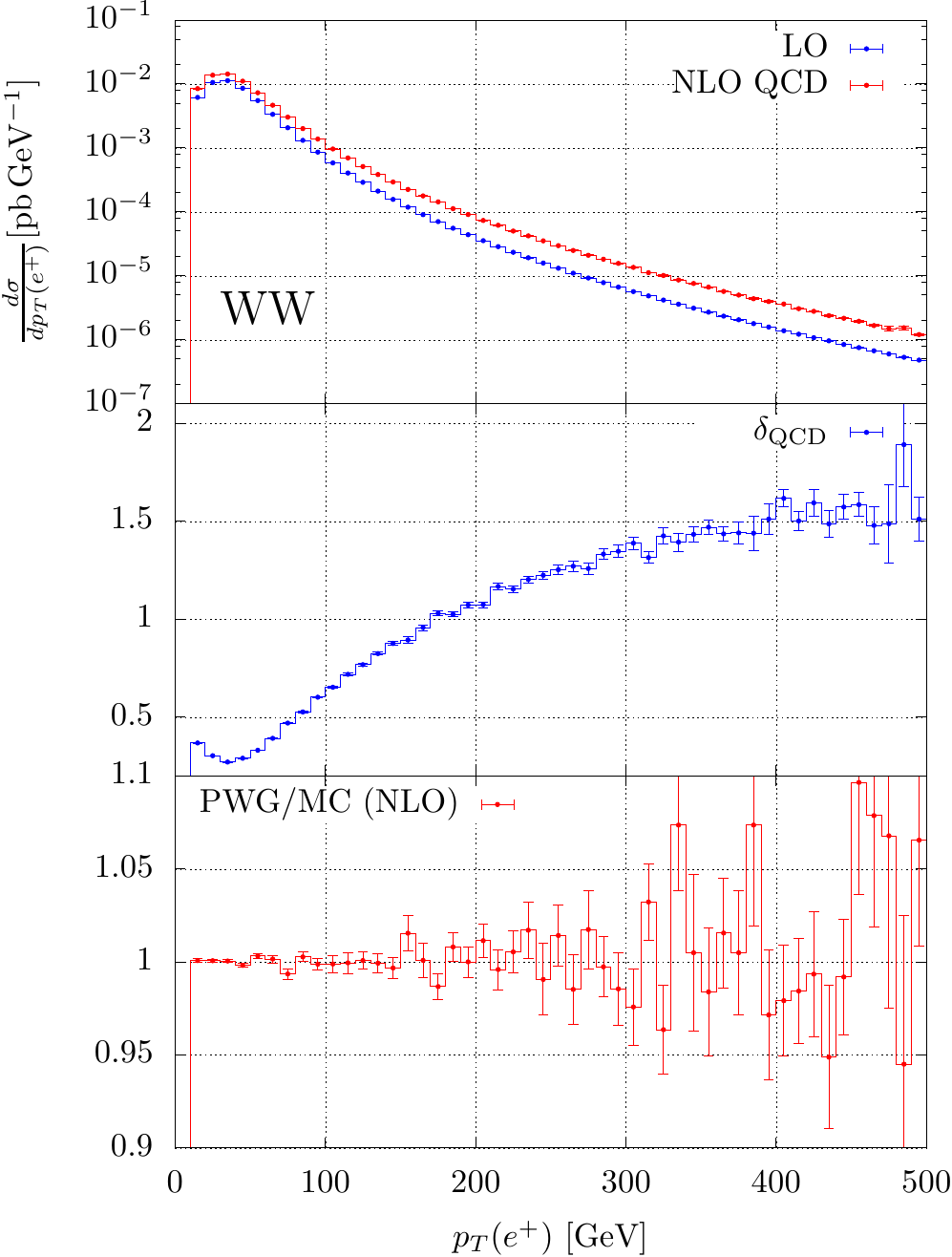}
    \includegraphics[scale=0.8]{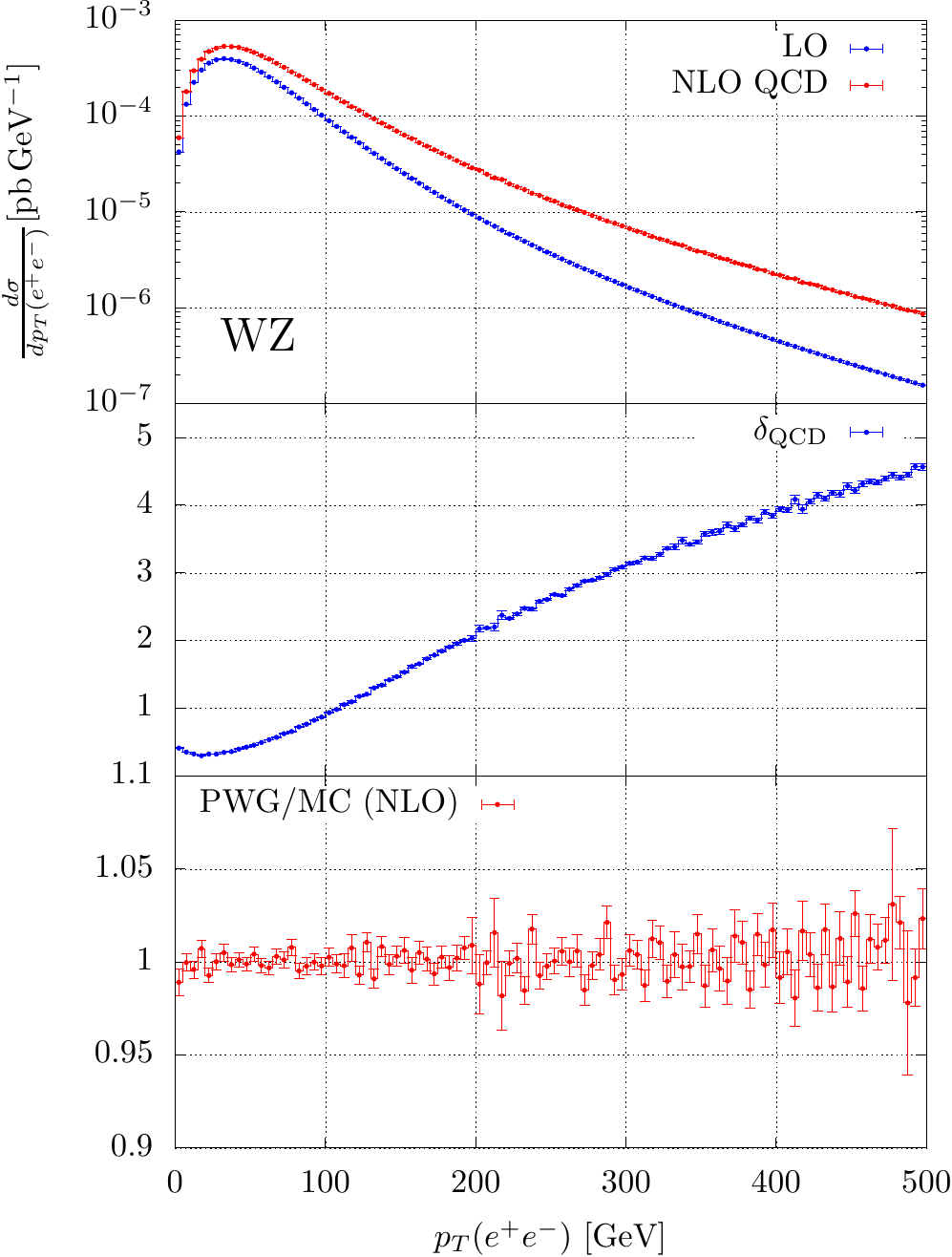}\\
    \includegraphics[scale=0.8]{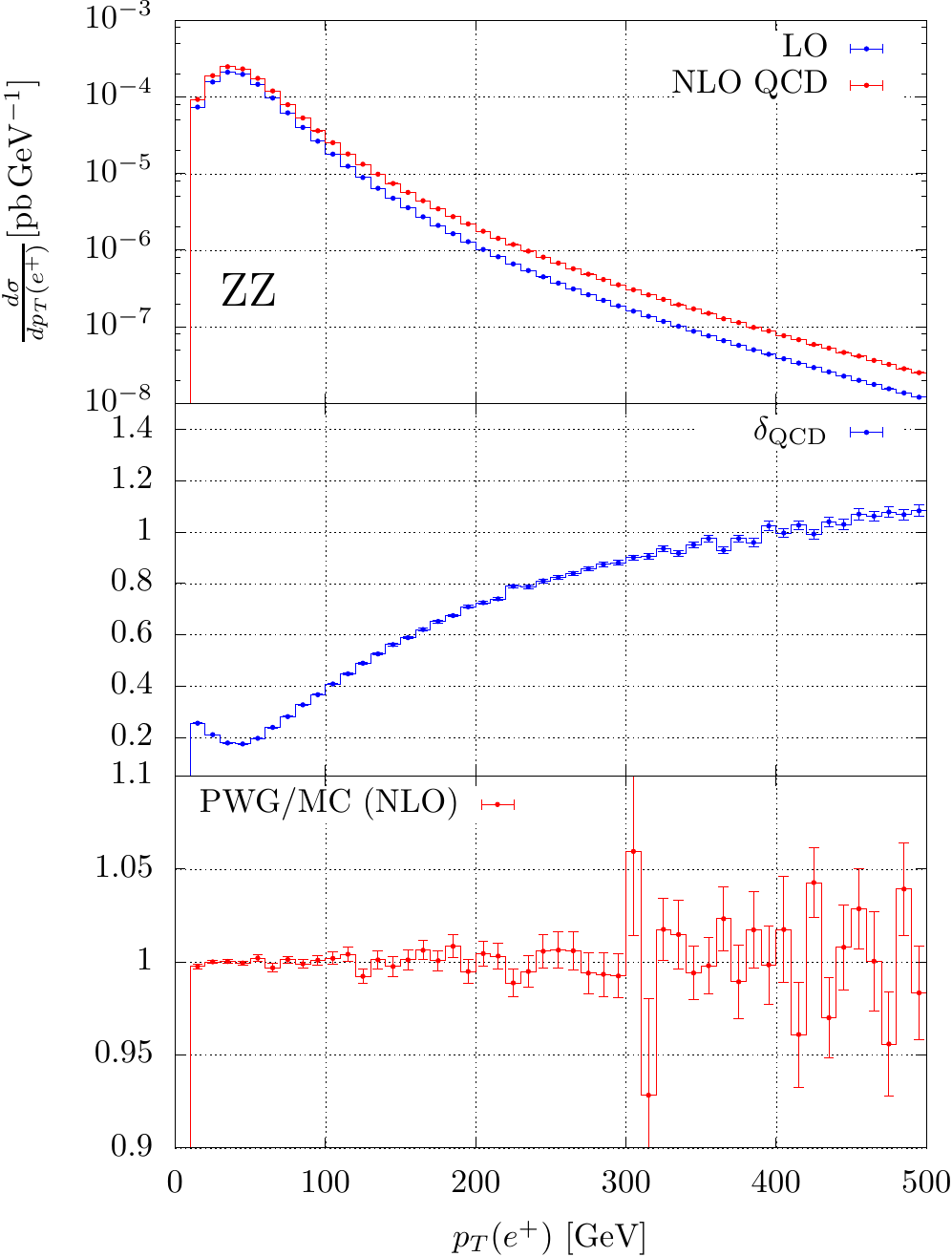}
    \includegraphics[scale=0.8]{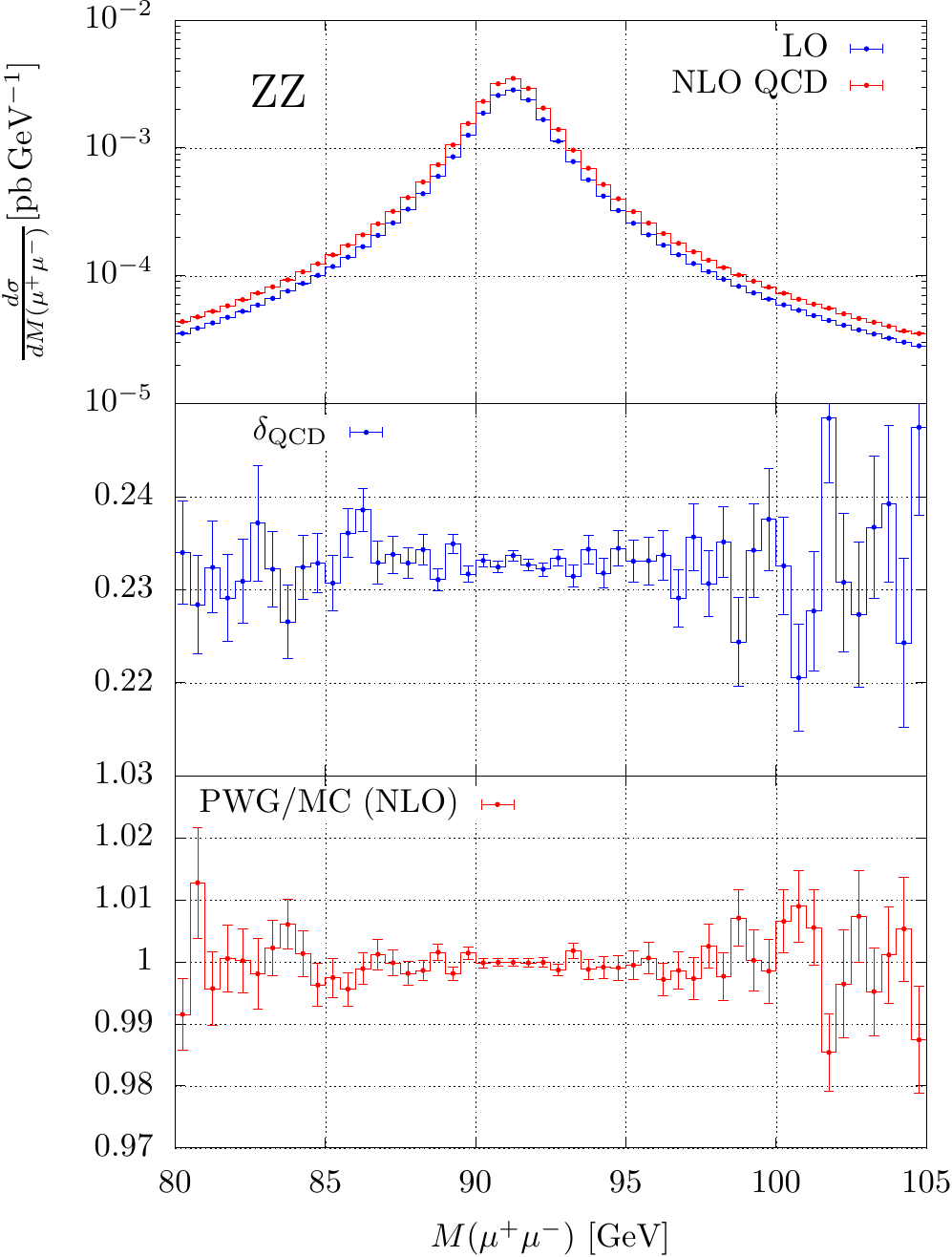}\\
    \caption{\label{fig:nloqcd} NLO QCD corrections to $pp \to e^+\nu_e \mu^-
      \overline{\nu}_\mu$ (top left), $pp \to \mu^+\nu_\mu e^- e^+$ (top
      right), and $pp \to \mu^+\mu^- e^- e^+$ (bottom) at the differential
      distribution level. Factorization and renormalization scales are set to
      the four-lepton invariant mass. Top panels: differential distributions
      at LO~(blue) and NLO~(red).  Central panels: relative NLO corrections
      ($\delta=$NLO/LO-1). Lower panels: ratio of the NLO QCD predictions
      computed with {\sc POWHEG} and {\sc MC}. See main text for details.}
  \end{center}
\end{figure*}

\begin{figure*}
  \begin{center}  
    \includegraphics[scale=0.8]{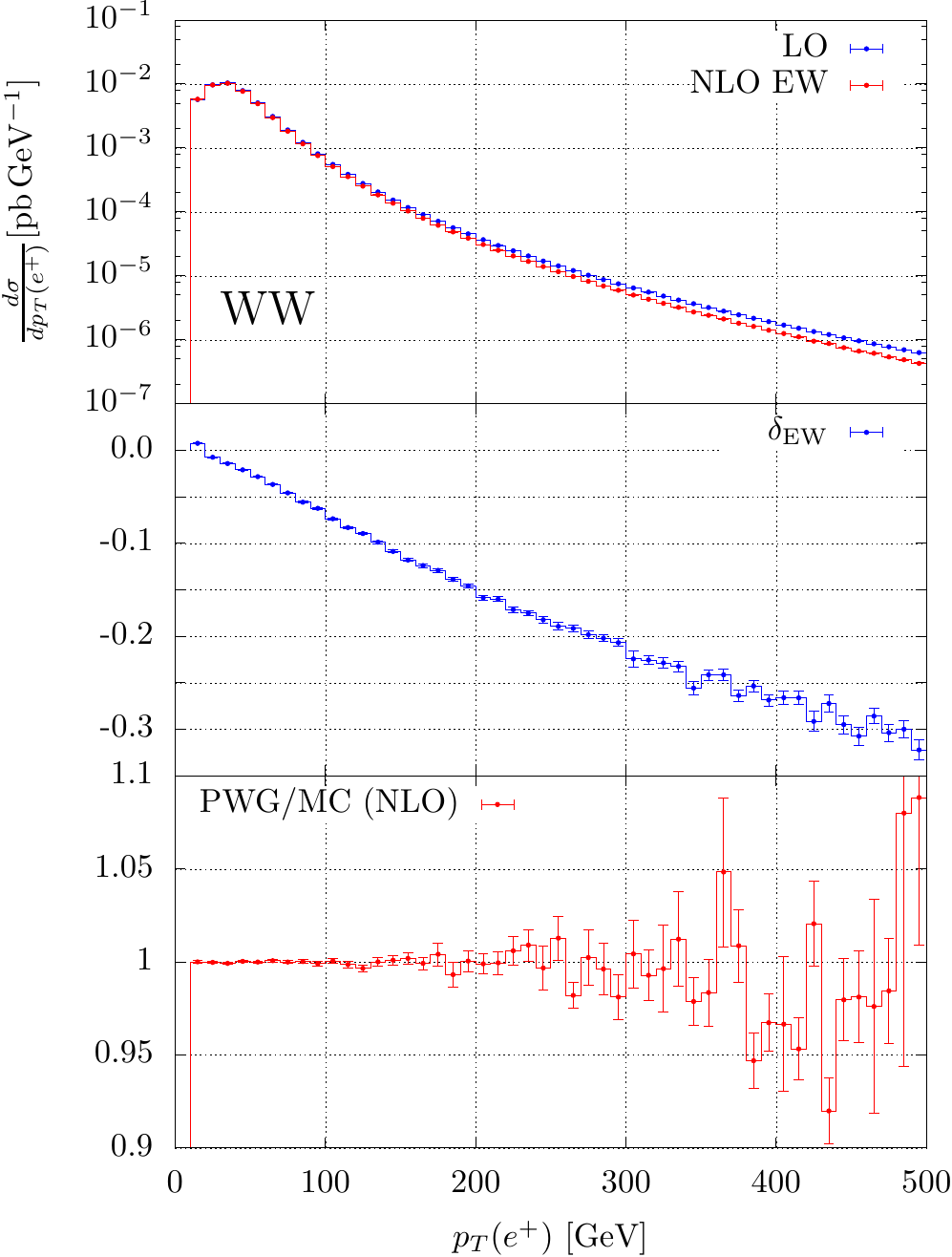}
    \includegraphics[scale=0.8]{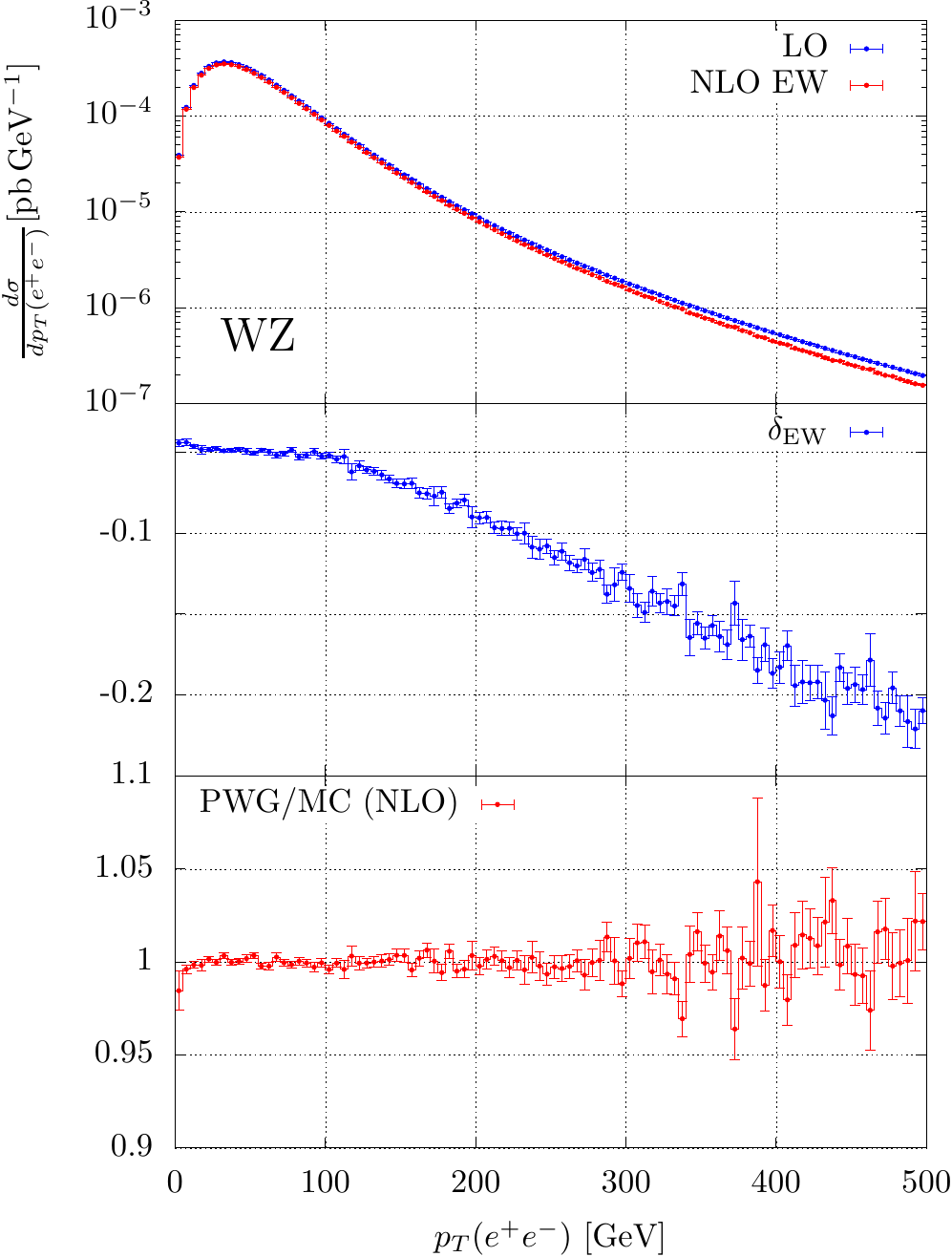}\\
    \includegraphics[scale=0.8]{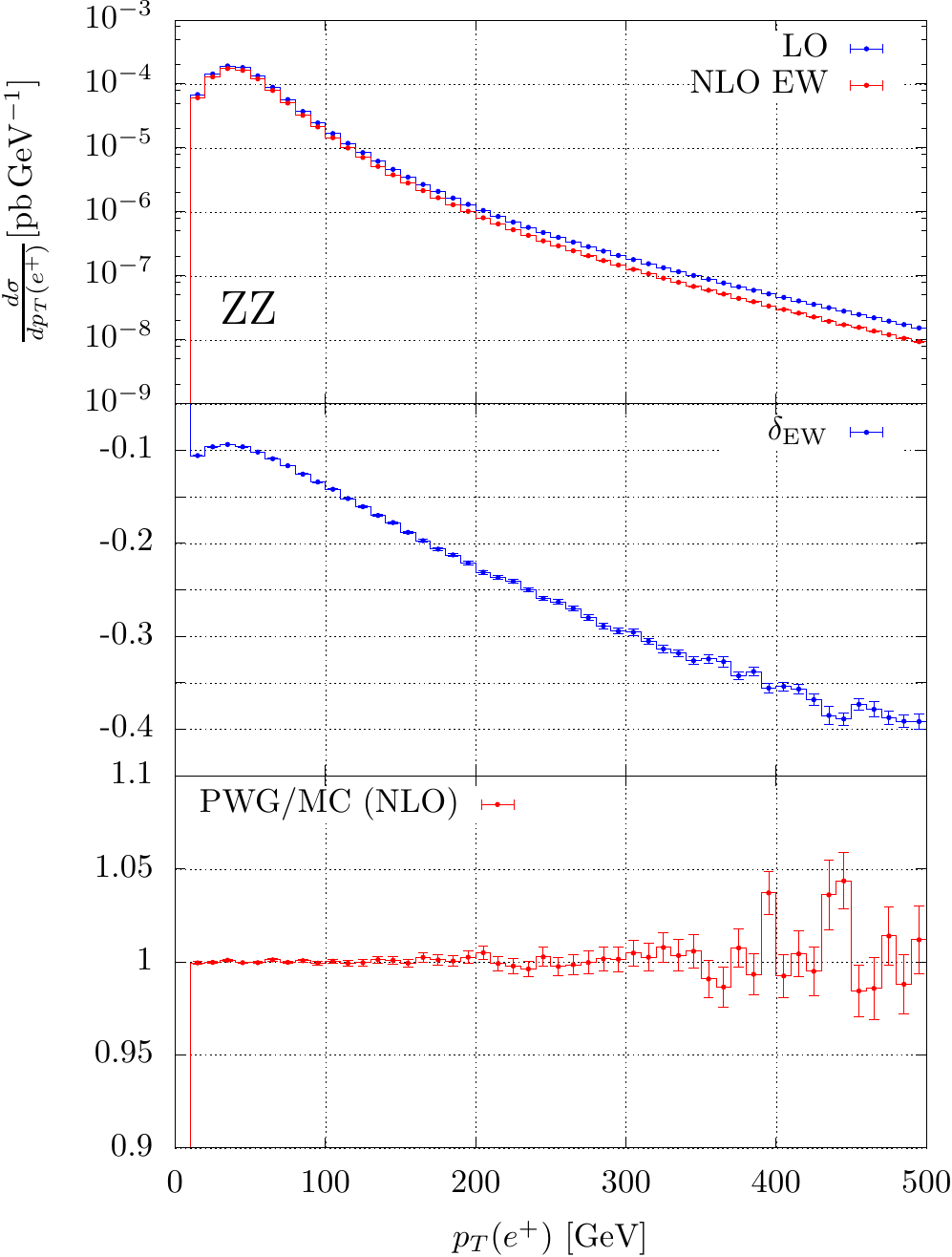}
    \includegraphics[scale=0.8]{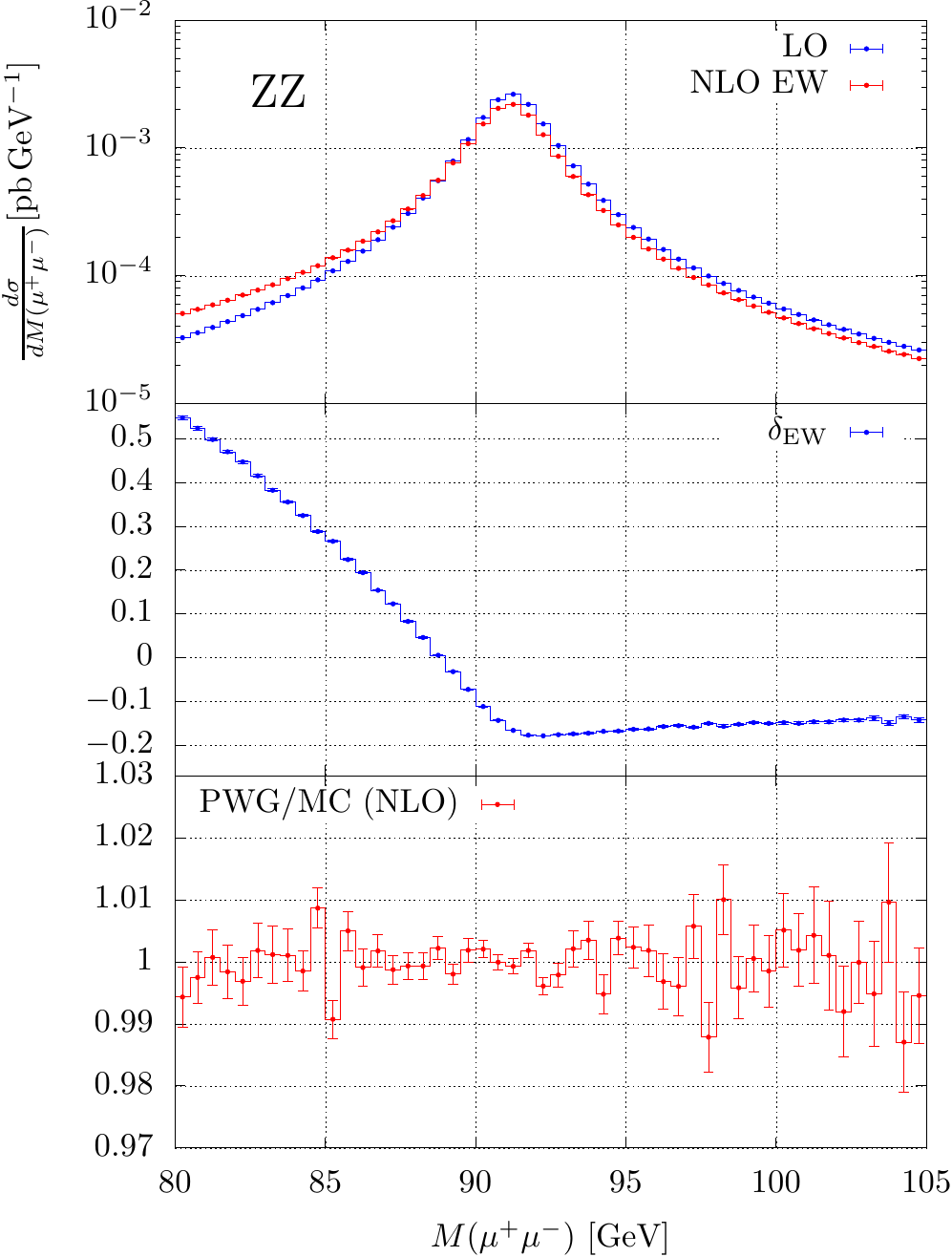}\\
    \caption{\label{fig:nloew} NLO EW corrections to $pp \to e^+\nu_e \mu^-
      \overline{\nu}_\mu$ (top left), $pp \to \mu^+\nu_\mu e^- e^+$ (top
      right), and $pp \to \mu^+\mu^- e^- e^+$ (bottom) at the differential
      distribution level. The factorization scale is set to $\mu=(M_V^{\rm
        \sss OS}+M_{V'}^{\rm \sss OS})/2$ ($V,V'=W,Z$).  Top panels:
      differential distributions at LO~(blue) and NLO~(red).  Central panels:
      relative NLO corrections ($\delta=$NLO/LO-1). Lower panels: ratio of
      the NLO EW predictions computed with {\sc POWHEG} and {\sc MC}. See
      main text for details.  }
  \end{center}
\end{figure*}

In order to validate the implementation of the $\nloqcd$ + $\nloew$
corrections to diboson-production processes in \RES{}, we compare the
predictions of our code with the ones of the Monte Carlo integrator used in
Ref.~\cite{Chiesa:2018lcs} ({\sc MC} in the following). Both codes use {\sc
  Recola2} for the calculation of the matrix elements, however, the
integration and the subtraction of the IR singularities is performed in a
completely independent way in the two programs.  In particular, {\sc POWHEG}
uses the FKS subtraction (modified to take into account the presence of
resonances), while in {\sc MC} the Catani-Seymour
procedure~\cite{Catani:1996vz, Dittmaier:1999mb} is used.

\allowbreak

Tables~\ref{tab:LOfix}-\ref{tab:NLOEW} collect the results at the integrated
cross-section level under the event selection of Eqs.~(\ref{eq:cuts})
and~(\ref{eq:cuts2}) for the processes $pp \to e^+\nu_e \mu^-
\overline{\nu}_\mu$, $pp \to \mu^+\nu_\mu e^- e^+$, and $pp \to \mu^+\mu^-
e^- e^+$ (dubbed ``WW'', ``WZ'', and ``ZZ'' in the tables).
Tables~\ref{tab:LOfix} and~\ref{tab:LOrun} show the results at LO for fixed
and running scales, Tabs.~\ref{tab:NLOQCDfix} and~\ref{tab:NLOQCDrun} contain
the results at $\nloqcd$ for fixed and running scales, while the predictions
at $\nloew$ accuracy (for fixed scales) are presented in
Tab.~\ref{tab:NLOEW}. The $\nloqcd$ corrections are positive, large and they
are dominated by real QCD corrections: this is a consequence of the opening
of gluon-induced channels ($qg\to VVq$) at $\nloqcd$.

Since we are focused on the technical comparison of the two programs, we do
not perform here scale-variation studies. The effect of scale variation can
be read, for instance, from Ref.~\cite{Chiesa:2018lcs} and turns out to be
large, given the size and the nature of the dominant contributions to the NLO
QCD corrections.

The NLO EW corrections are negative and moderate at the integrated
cross-section level. They are a combination of QED and purely weak effects.

In Tabs.~\ref{tab:LOfix}-\ref{tab:NLOEW}, the numbers in parenthesis
correspond to the statistical integration error on the last digit. As can be
seen from Tabs.~\ref{tab:LOfix}-\ref{tab:NLOEW}, the predictions of the two
programs agree within the integration error.

The comparison at the differential distribution level is shown in
Figs.~\ref{fig:nloqcd} and~\ref{fig:nloew} for the NLO QCD (running scales)
and the NLO EW (fixed scales) predictions, respectively. For $WW$ and $ZZ$
production we consider the transverse momentum of the positron, $\pT(e^+)$,
while for $WZ$ production we take the transverse momentum of the $e^+e^-$
pair, $\pT(e^+e^-)$, i.e.~the reconstructed $Z$. For $ZZ$ production, we also
present the results for the invariant mass of the $\mu^+\mu^-$ pair,
$M(\mu^+\mu^-)$.  In Figs.~\ref{fig:nloqcd} and~\ref{fig:nloew}, the upper
panels show the differential distributions at LO~(blue) and NLO~(red)
accuracy, the central panels show the relative NLO corrections ($\delta={\rm
  NLO}/{\rm LO}-1$), while, in the lower panels, we plot the ratio of the NLO
predictions from {\sc POWHEG} and {\sc MC}. As largely discussed in the
literature, the NLO QCD corrections to the transverse momentum observables
are positive, large, and increase with $\pT$. On the contrary, the NLO QCD
corrections to $M(\mu^+\mu^-)$ are flat and correspond to a normalization
factor. The NLO EW corrections to the transverse-momentum distributions are
negative and show the typical Sudakov behaviour~\cite{Beenakker:1993tt,
  Beccaria:1998qe, Ciafaloni:1998xg, Kuhn:1999de, Ciafaloni:2000df,
  Denner:2000jv, Denner:2001gw} at high $\pT$. The shape of the NLO EW
corrections to $M(\mu^+\mu^-)$ is dominated by QED effects, since the
radiation of a final-state photon reduces the invariant mass of the lepton
pair, shifting the events from the peak of the LO distribution to the region
below the $Z$ resonance.  As in the case of the cross-section level
comparison, from the lower panels of Figs.~\ref{fig:nloqcd}
and~\ref{fig:nloew} we conclude that {\sc POWHEG} and {\sc MC} agree within
the statistical errors.

For the fixed-order part of the calculation, {\sc POWHEG} computes the
$\nloqcd$ + $\nloew$ corrections additively. We checked that, when running
our code with both QCD and EW corrections, $\delta_{\rm QCD+EW}$ from {\sc
  POWHEG} is equal to the sum of $\delta_{\rm QCD}$ and $\delta_{\rm EW}$
computed with {\sc MC}. We do not show here the plots, since the
corresponding information can be read from the combination of
Figs.~\ref{fig:nloqcd} and~\ref{fig:nloew}.

\section{Results at NLO+PS accuracy}
\label{sec:nlops}

\begin{figure*}
  \begin{center}  
    \includegraphics[scale=0.8]{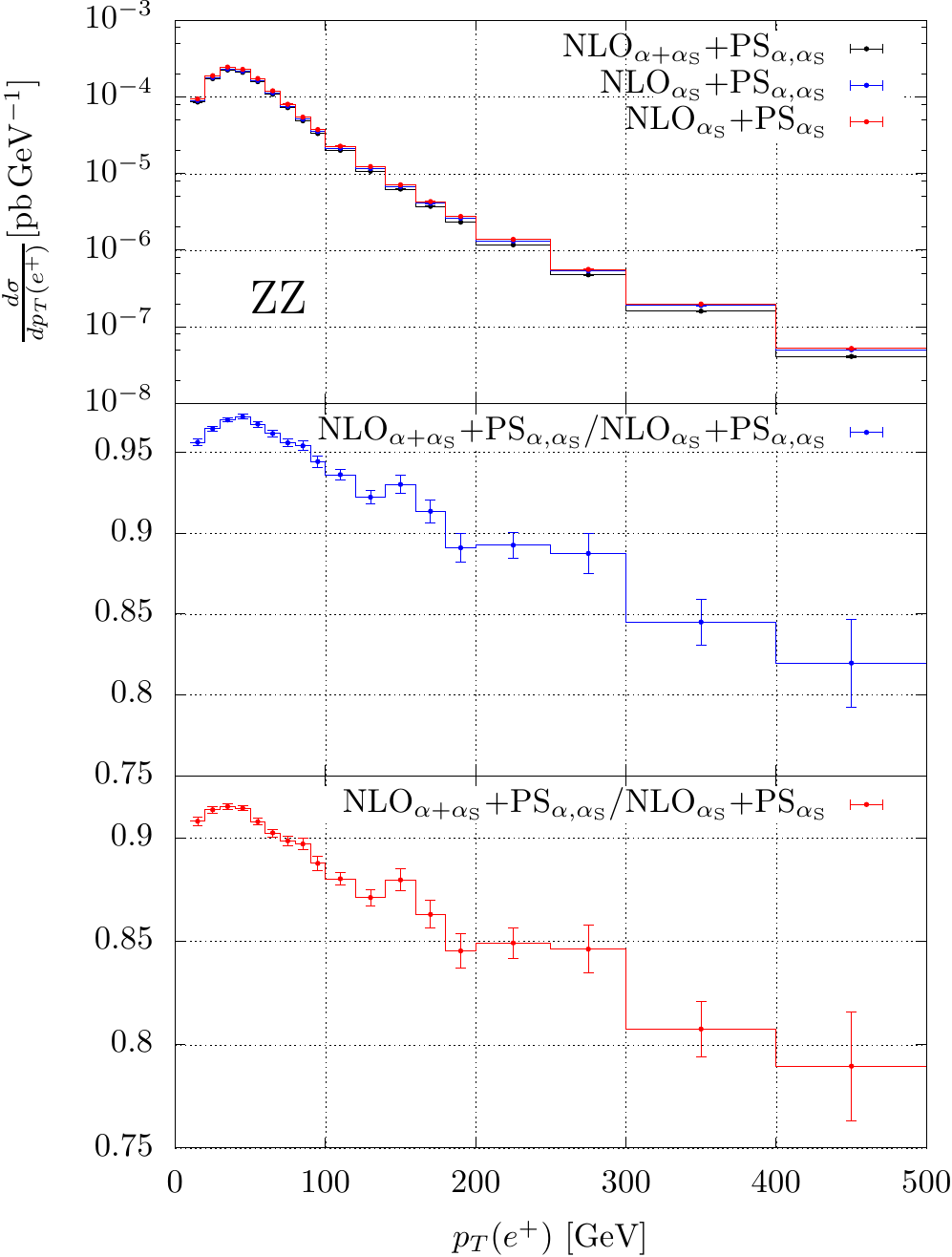}
    \includegraphics[scale=0.8]{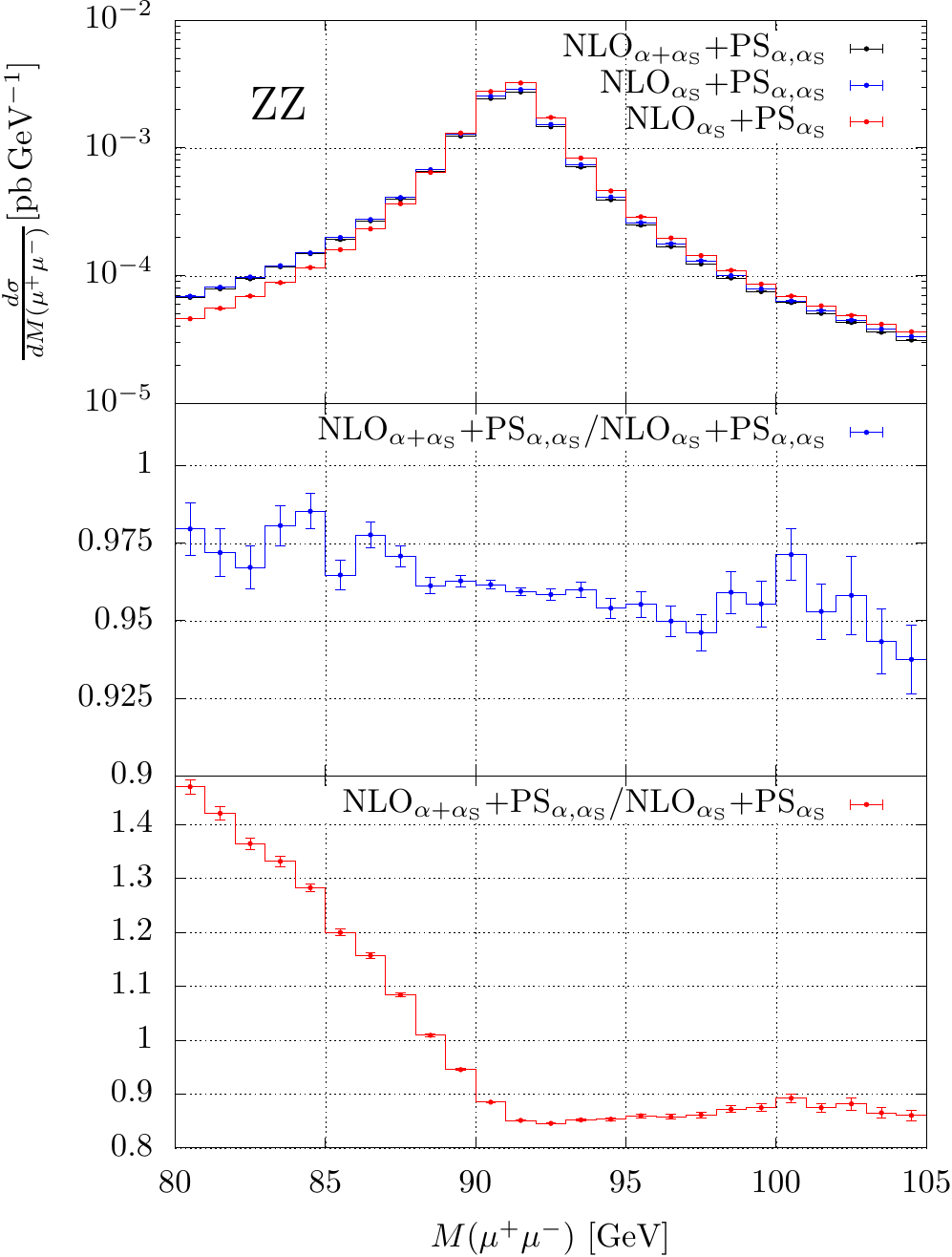}\\
    \includegraphics[scale=0.8]{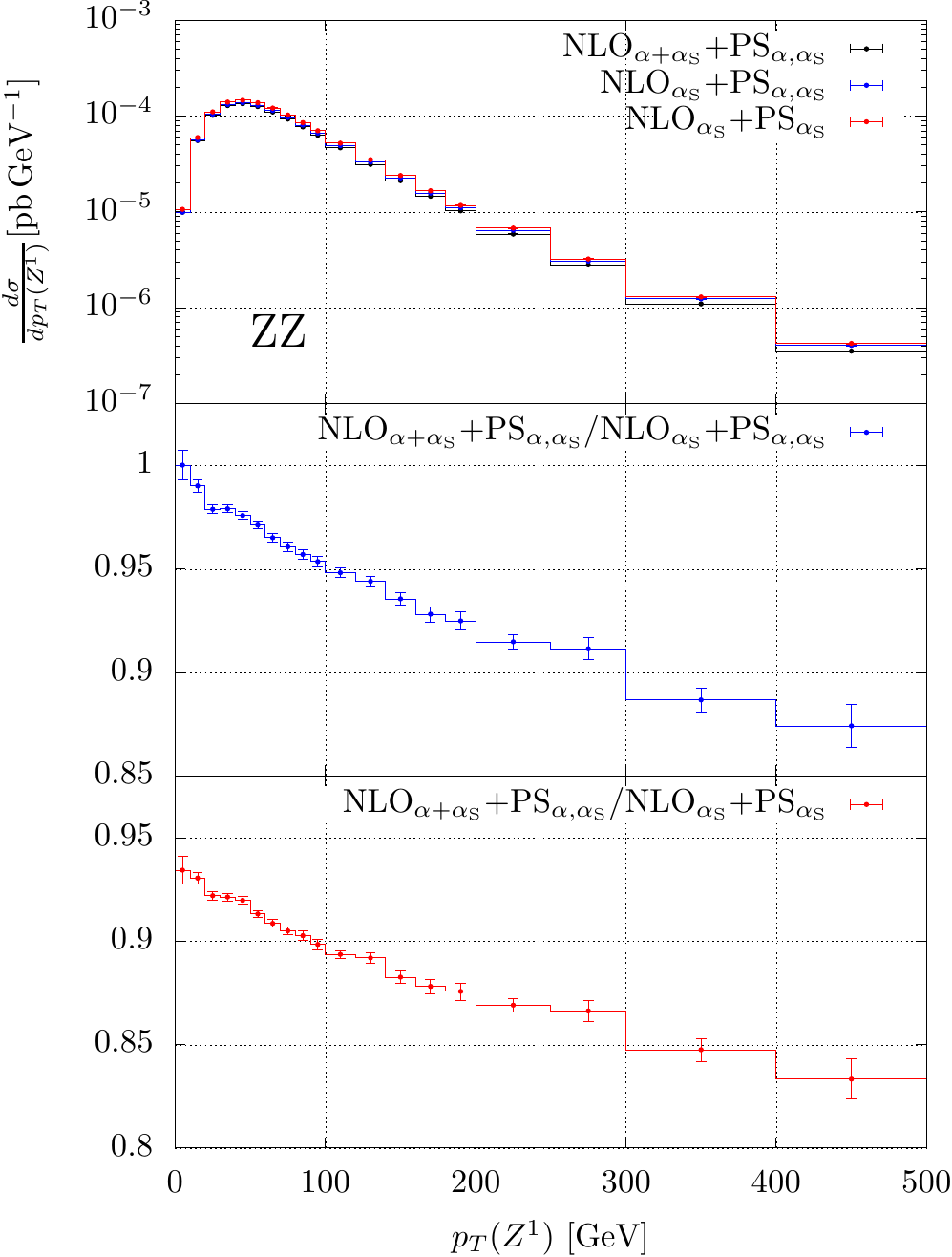}
    \includegraphics[scale=0.8]{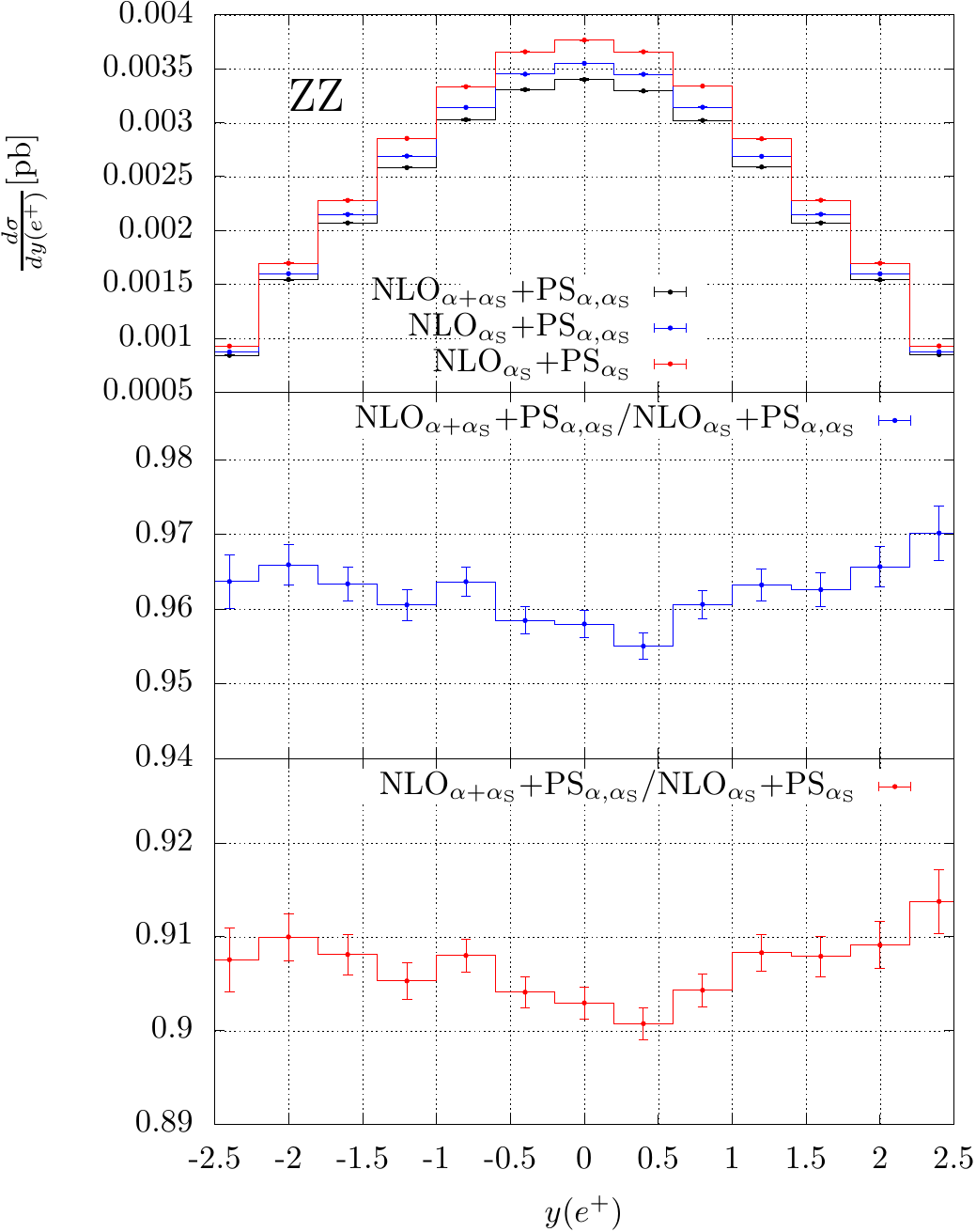}\\
    \caption{\label{fig:nlopszz} Comparison of the predictions at $\nloqcd$ +
      $\nloew$ + $\qcdqedps$ (NLO$_{\alpha+\alpha_{\rm S}}$+
      PS$_{\alpha,\alpha_{\rm S}}$), at $\nloqcd$ + $\qcdqedps$
      (NLO$_{\alpha_{\rm S}}$+PS$_{\alpha,\alpha_{\rm S}}$), and at $\nloqcd$
      + $\qcdps$ (NLO$_{\alpha_{\rm S}}$ + PS$_{\alpha_{\rm S}}$) accuracy
      for the process $pp \to \mu^+\mu^- e^- e^+$.  Upper panels:
      differential distributions as a function of the positron transverse
      momentum (top left), of the dimuon invariant mass (top right), of the
      transverse momentum of the hardest $Z$ (bottom left), and of the
      positron rapidity (bottom right).  Central panels: ratio of the
      predictions at $\nloqcd$ + $\nloew$ + $\qcdqedps$ and at $\nloqcd$ +
      $\qcdqedps$. Lower panels: ratio of the results at $\nloqcd$ + $\nloew$
      + $\qcdqedps$ and at $\nloqcd$ + $\qcdps$. See main text for details.}
  \end{center}
\end{figure*}

\begin{figure*}
  \begin{center}  
    \includegraphics[scale=0.8]{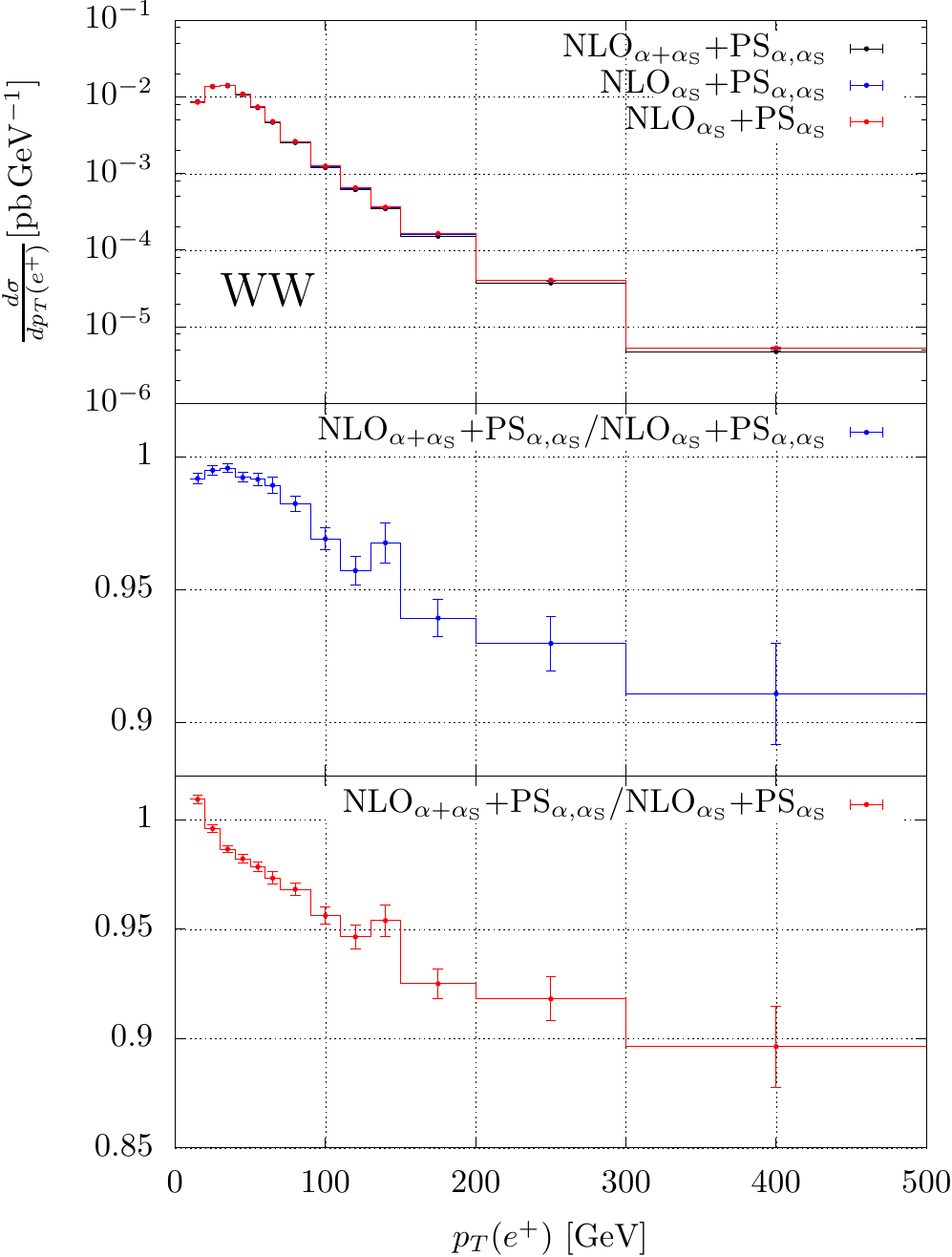}
    \includegraphics[scale=0.8]{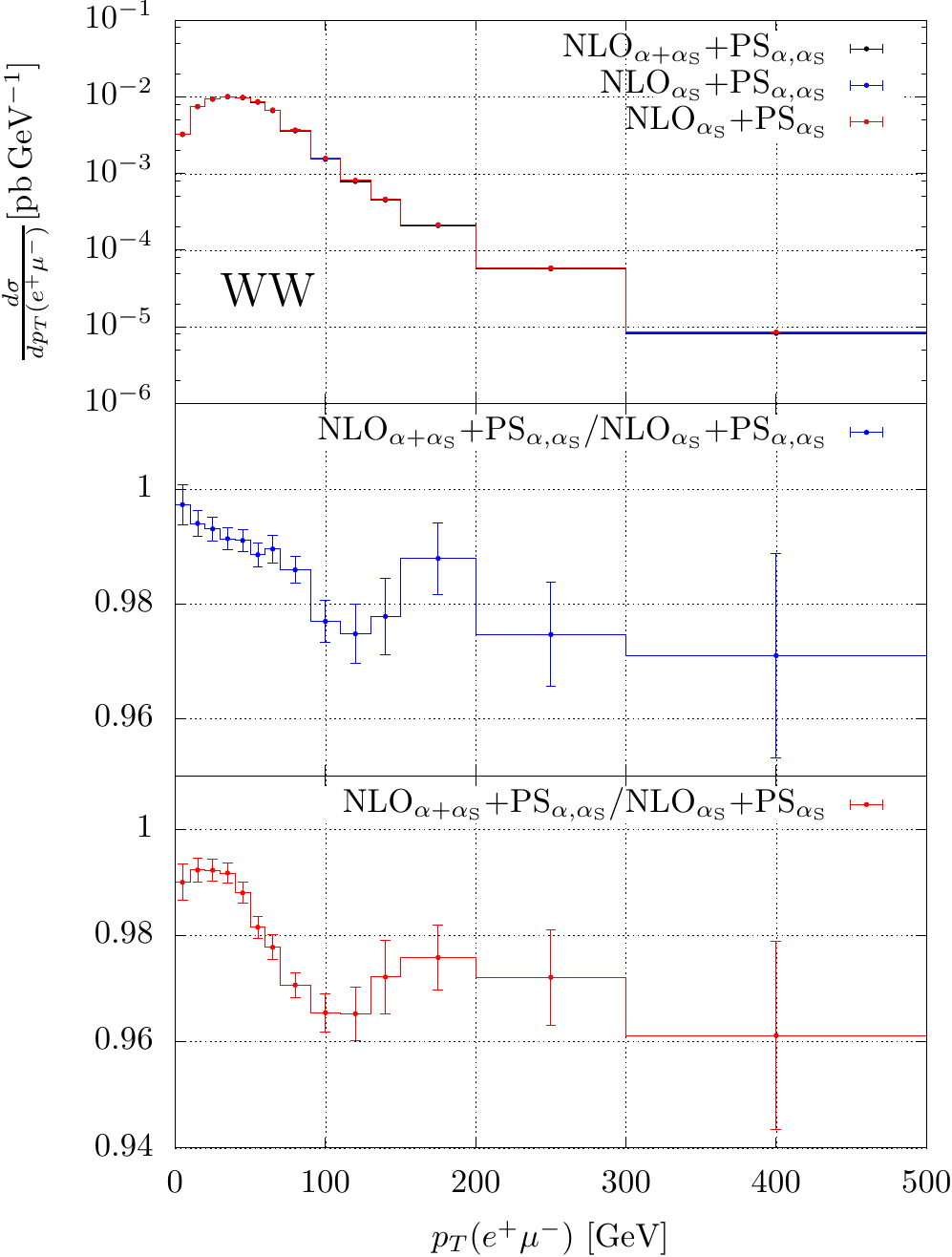}\\
    \includegraphics[scale=0.8]{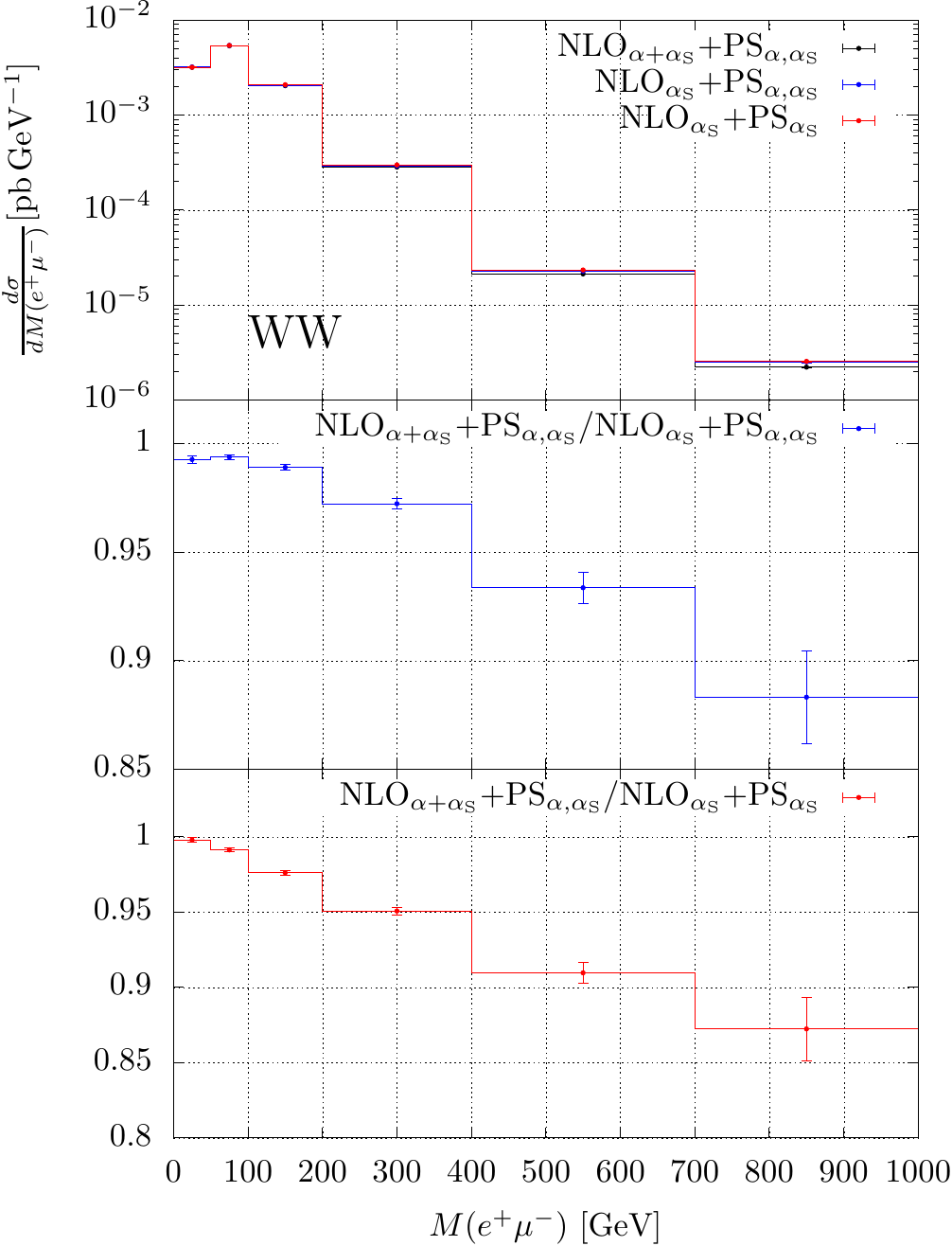}
    \includegraphics[scale=0.8]{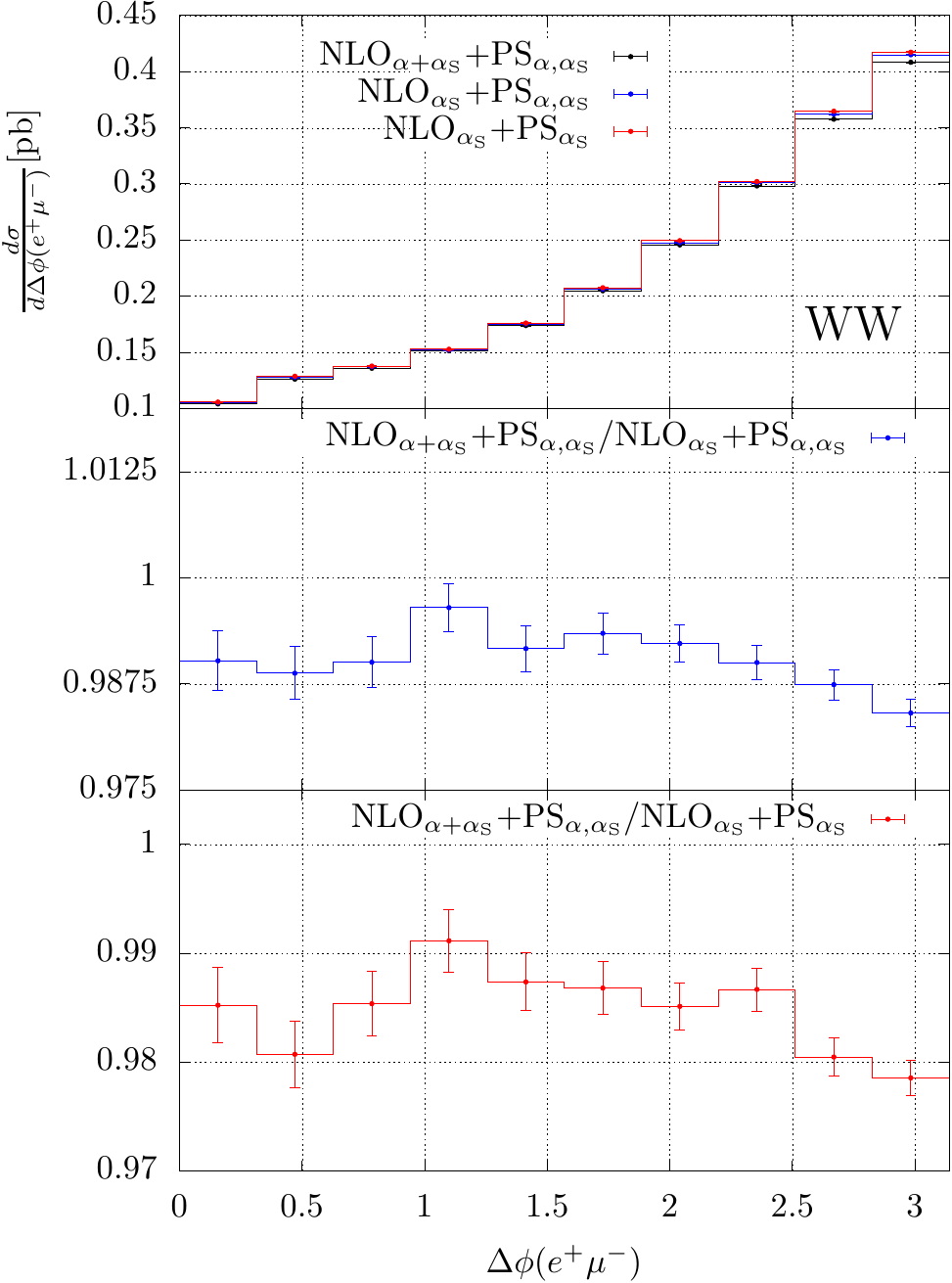}\\
    \caption{\label{fig:nlopsww} Comparison of the predictions at $\nloqcd$ +
      $\nloew$ + $\qcdqedps$ (NLO$_{\alpha+\alpha_{\rm S}}$+
      PS$_{\alpha,\alpha_{\rm S}}$), at $\nloqcd$ + $\qcdqedps$
      (NLO$_{\alpha_{\rm S}}$+PS$_{\alpha,\alpha_{\rm S}}$), and at $\nloqcd$
      + $\qcdps$ (NLO$_{\alpha_{\rm S}}$ + PS$_{\alpha_{\rm S}}$) accuracy
      for the process $pp \to e^+\nu_e \mu^-\overline{\nu}_\mu$.  Upper
      panels: differential distributions as a function of the positron
      transverse momentum (top left), of the transverse momentum (top right)
      and invariant mass (bottom left) of the positron--muon system, and of
      the azimuthal-angle separation between the positron and the muon
      (bottom right).  Central panels: ratio of the predictions at $\nloqcd$
      + $\nloew$ + $\qcdqedps$ and at $\nloqcd$ + $\qcdqedps$. Lower panels:
      ratio of the results at $\nloqcd$ + $\nloew$ + $\qcdqedps$ and at
      $\nloqcd$ + $\qcdps$. See main text for details.}
  \end{center}
\end{figure*}

In this section, we present the results at NLO accuracy matched to PS. For
brevity, we only show results for $ZZ$ and $WW$ production, but the code can
be used to generate events and perform a similar study for $W^\pm Z$, as
well. We consider three different levels of accuracy:
\begin{itemize}
\item $\nloqcd$ + $\nloew$ + $\qcdqedps$: full NLO corrections matched to the
  full PS with QED and QCD radiation (NLO$_{\alpha+\alpha_{\rm S}}$ +
  PS$_{\alpha,\alpha_{\rm S}}$ in the plots);
\item $\nloqcd$ + $\qcdqedps$: strong corrections matched to the full PS
  (NLO$_{\alpha_{\rm S}}$ + PS$_{\alpha,\alpha_{\rm S}}$ in the plots);
\item $\nloqcd$ + $\qcdps$: strong corrections matched to a PS without QED
  radiation (NLO$_{\alpha_{\rm S}}$ + PS$_{\alpha_{\rm S}}$ in the plots).
\end{itemize}
For the predictions at $\nloqcd$ + $\nloew$ + $\qcdqedps$, according to the
{\tt allrad} scheme, our code generates up to three emissions, namely ISR QCD
or QED radiation, and FSR QED radiation from the decay products of each one
of the vector bosons.
The kinematics of the hard partonic event generated by {\sc POWHEG}, together
with the values of the transverse momenta (with respect to their emitters) of
the generated partonic and/or photonic radiation, is then saved in the Les
Houches event file.  The transverse momentum of the initial-state radiation,
if present, is used by the parton shower algorithm as upper bound for the
generation of QED/QCD radiation from the hard production process.  The
transverse momentum of the photons from the final-state leptons (i.e.~from
the resonances) is used by the parton shower program as upper bound for
further QED radiation.
The results presented in this paper have been showered by {\sc PYTHIA8}. This
code allows to veto emissions harder than the ones generated by {\sc POWHEG}
by using dedicated {\tt UserHooks}. We have also verified that we obtain
fully compatible results if we let the PS generate unconstrained emissions
and we subsequently check if the transverse momentum of the radiations with
respect to the emitting particles is smaller than the {\sc POWHEG} hardest
ones. If this is not the case, we attempt to shower the event again until all
constraints are met and the event is accepted.

For the predictions at $\nloqcd$ + $\qcdqedps$, the LH events contain at most
one initial-state QCD radiation and the transverse momentum of the radiated
parton sets the maximum hardness for the QCD PS, while the starting scale for
the QED PS is the center of mass energy of the event for ISR, and the
virtuality of the resonances for FSR.  The predictions at $\nloqcd$ matched
to QCD PS are obtained from the same LH events used for the study at
$\nloqcd$ + $\qcdqedps$ accuracy, simply by turning off the QED radiation in
{\sc PYTHIA}.

In Figs.~\ref{fig:nlopszz} and~\ref{fig:nlopsww}, the upper panels show the
differential cross section as a function of the observable under
consideration, the central panels contain the ratio of the results at
$\nloqcd$ + $\nloew$ + $\qcdqedps$ and the ones at $\nloqcd$ + $\qcdqedps$,
while, in the lower panels, we show the ratio of the predictions at $\nloqcd$
+ $\nloew$ + $\qcdqedps$ and the ones at $\nloqcd$ + $\qcdps$.

The calculation at $\nloqcd$ + $\nloew$ accuracy matched to the full PS
($\qcdqedps$) includes the effect of the one-loop virtual weak corrections,
the full QED corrections at $\mathcal{O}(\alpha)$,\footnote{Strictly
  speaking, a gauge invariant separation of QED and weak effects beyond
  leading logarithmic accuracy is only possible for $ZZ$ production.} and
part of the mixed factorized corrections at $\mathcal{O}(\alpha\alpha_{\rm
  S})$ (coming from the product of the NLO normalization encoded in the {\sc
  POWHEG} $\bar{B}$ function and the Sudakov form factors in the {\sc POWHEG}
master formula for event generation~\cite{Alioli:2010xd}). Therefore the
central panels of Figs.~\ref{fig:nlopszz} and~\ref{fig:nlopsww} show the
effect of the weak NLO corrections, the difference between the exact QED
corrections at $\mathcal{O}(\alpha)$ and their PS approximation, and the
mixed corrections.

In the lower panels, the ratios are taken with respect to a result where only
QCD corrections are included. Therefore, on top of the same effects as in the
central panels, these panels also include the effect of all-order photonic
corrections (without approximations at $\mathcal{O}(\alpha)$, and in PS
approximation starting from $\mathcal{O}(\alpha^2)$).

For the process $pp \to \mu^+\mu^- e^- e^+$ ($ZZ$ production,
Fig.~\ref{fig:nlopszz}), besides the observables used in
Sect.~\ref{sec:checks} for the validation at NLO ($\pT(e^+)$ and
$M(\mu^+\mu^-)$), we consider the transverse momentum of the hardest
reconstructed $Z$ boson, $\pT(Z^1)$, and the positron rapidity, $y(e^+)$. For
the transverse-momentum distributions (left plots), the predictions at
$\nloqcd$ + $\nloew$ + $\qcdqedps$ are always lower than the ones at
$\nloqcd$ + $\qcdqedps$, or at $\nloqcd$ + $\qcdps$, in particular at high
$\pT$, where the EW Sudakov corrections amount to approximately $-30$\% with
respect to the LO. The photonic corrections further reduce the predictions.
The ratios for the positron rapidity distribution (bottom right plot) are
essentially flat and, in the central panel, show an effect of about $-3/-4$\%
mainly coming from weak corrections that becomes approximately $-10$\% in the
lower panel, where the denominator does not include photonic corrections. The
ratio of the predictions at $\nloqcd$ + $\nloew$ + $\qcdqedps$ and the ones
at $\nloqcd$ + $\qcdqedps$ for the dimuon invariant-mass distribution (top
right plot) is rather flat and shows that the effect of the EW corrections
beyond QED PS amounts to about $-4$\%.  In the lower panel, there is a
positive corrections below the $Z$ peak coming from multiple photon radiation
(radiative return) very similar to the one observed at fixed order in
Fig.~\ref{fig:nloew}.

The predictions at NLO+PS accuracy for the process $pp \to e^+\nu_e \mu^-
\overline{\nu}_\mu$~($WW$) are collected in Fig.~\ref{fig:nlopsww} as a
function of the following observables: transverse momentum of the positron,
$\pT(e^+)$, transverse momentum and invariant mass of the positron--muon
system, $\pT(e^+\mu^-)$ and $M(e^+\mu^-)$, and azimuthal distance between the
positron and the muon, $\Delta \phi (e^+\mu^-)$. For all the observables
under consideration, the inclusion of the NLO EW corrections lowers the
predictions with respect to the calculation at $\nloqcd$ + $\qcdqedps$
(central panels). This effect is more pronounced in the tails of the
transverse-momentum and invariant-mass distributions, where the NLO EW
corrections are negative and large because of the EW Sudakov logarithms. A
comment is in order concerning the $\pt(e^+\mu^-)$ distribution.  In the
high-$\pt$ tail, the errors are large and prevent a clean assessment of the
size of the EW contributions. This is due to the fact that, in this region,
the QCD corrections are positive, large, and increase very steeply starting
from about 100~GeV in the absence of a jet veto.\footnote{This effect is
  similar to the one discussed in Ref.~\cite{Rubin:2010xp}, and it is related
  to the kinematic configuration where a vector boson recoils against a hard
  jet, and the remaining vector boson is relatively soft.}
From the lower panels of Fig.~\ref{fig:nlopsww}, we conclude that the
contribution of multiple photon radiation to the observables under
consideration (with the event selection of Eq.~(\ref{eq:cuts})) is negative
with the only exception of the first few bins of the $\pT(e^+)$ and
$M(e^+\mu^-)$ distributions.

\begin{figure*}
  \begin{center}
    \includegraphics[scale=0.8]{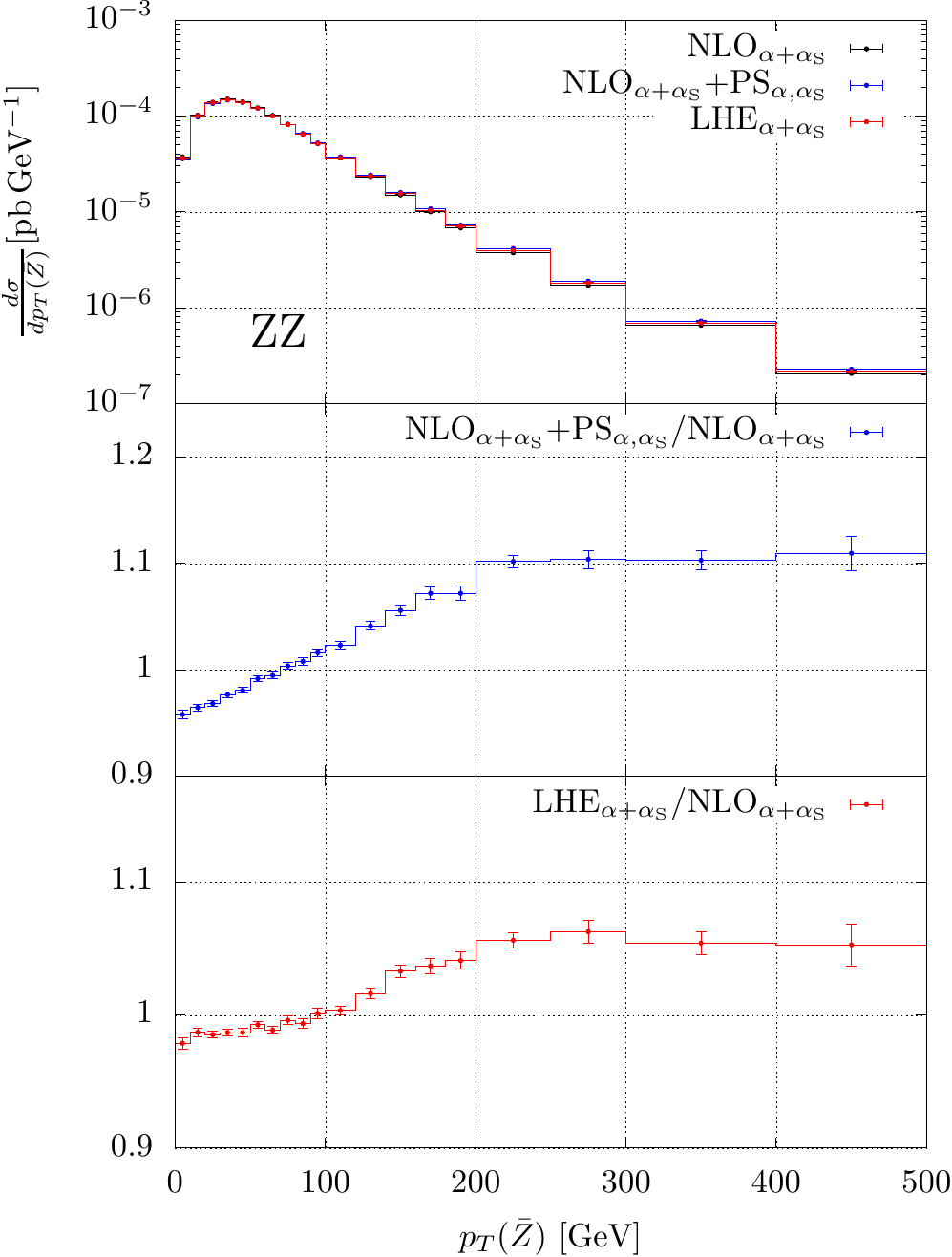}
    \includegraphics[scale=0.8]{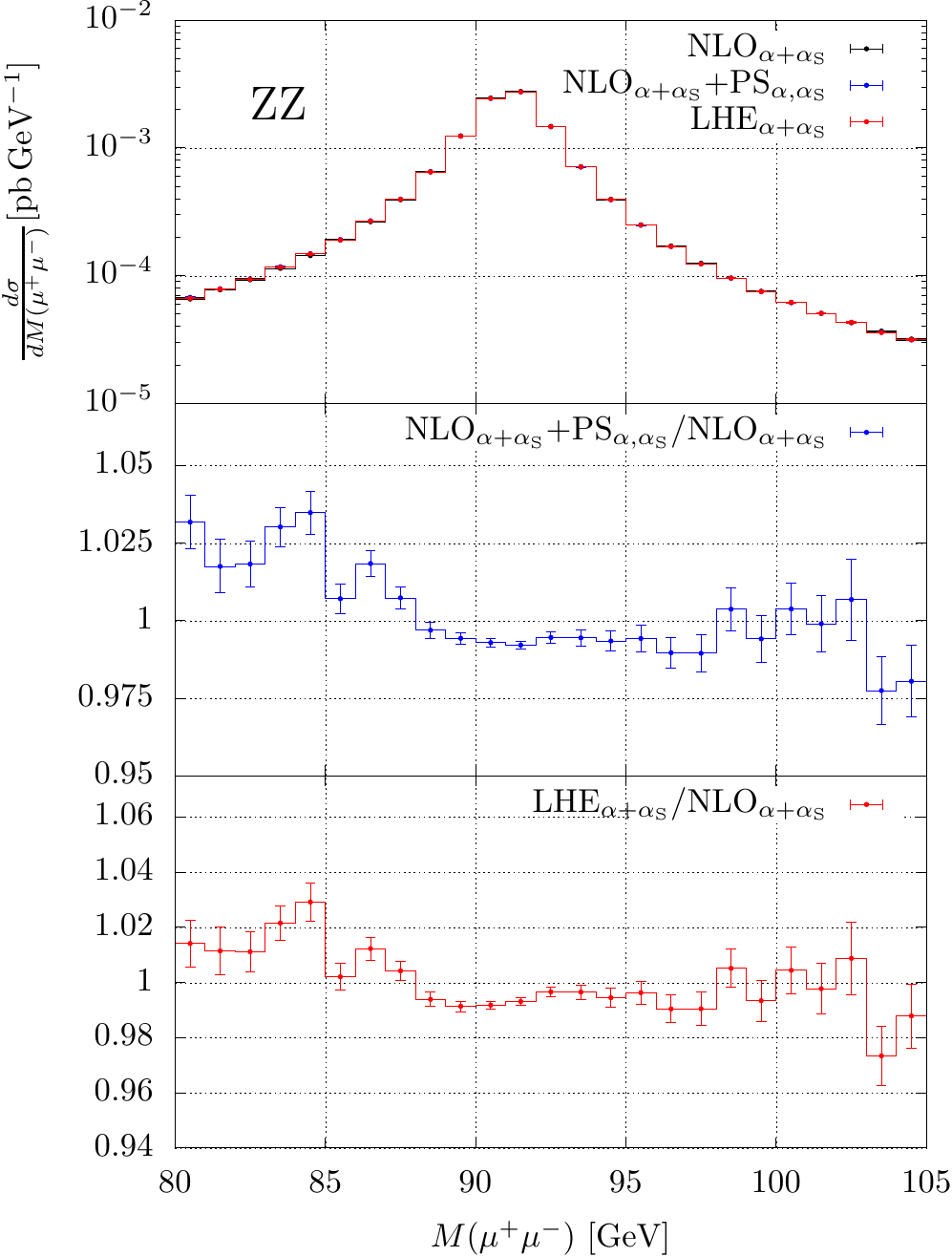}\\
    \includegraphics[scale=0.8]{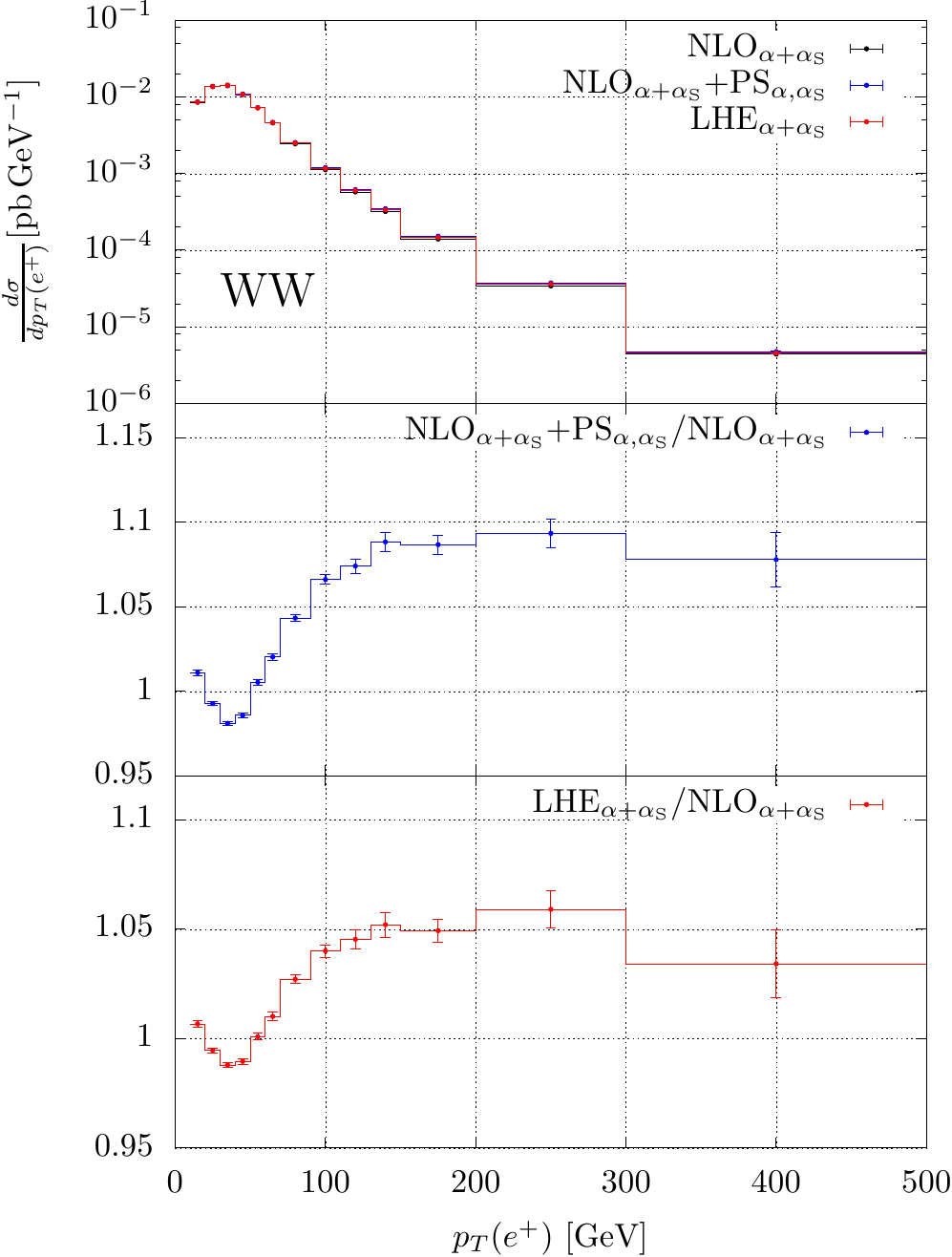}
    \includegraphics[scale=0.8]{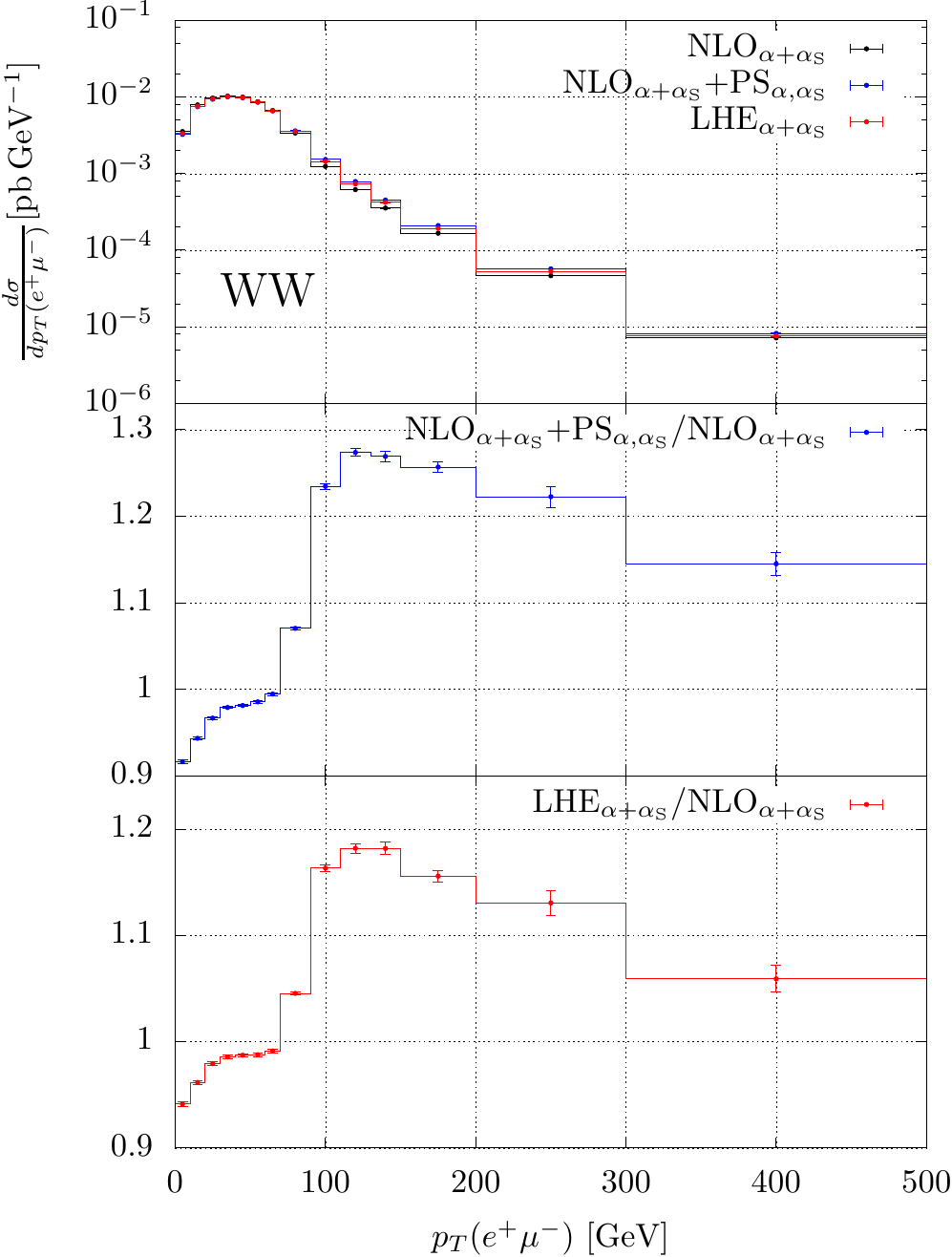}\\
    \caption{\label{fig:nlopslhe} Comparison of the predictions at $\nloqcd$
      + $\nloew$ + $\qcdqedps$ (NLO$_{\alpha+\alpha_{\rm S}}$+
      PS$_{\alpha,\alpha_{\rm S}}$), at LHE (LHE$_{\alpha,\alpha_{\rm S}}$),
      and at fixed-order $\nloqcd$ + $\nloew$ (NLO$_{\alpha+\alpha_{\rm S}}$)
      accuracy for the processes $pp \to \mu^+\mu^- e^- e^+$ (upper plots)
      and $pp \to e^+\nu_e \mu^-\overline{\nu}_\mu$ (lower plots).  Upper
      panels: differential distributions as a function of the transverse
      momentum of the same-flavour lepton pair, closer in virtuality to the
      nominal $Z$ mass (top left), of the dimuon invariant mass (top right),
      and of the transverse momentum of the positron (bottom left) and of the
      positron--muon system (bottom right).  Central panels: ratio of the
      predictions at $\nloqcd$ + $\nloew$ + $\qcdqedps$ and at fixed-order
      $\nloqcd$ + $\nloew$ accuracy.  Lower panels: ratio of the results at
      the LHE level and at fixed-order $\nloqcd$ + $\nloew$ accuracy.  See
      main text for details.  }
  \end{center}
\end{figure*}

In order to further discuss effects beyond fixed order, in addition to those
encountered so far, in Fig.~\ref{fig:nlopslhe} we show a few kinematic
distributions at $\nloqcd$ + $\nloew$ (NLO$_{\alpha+\alpha_{\rm S}}$ in the
plots, black lines), $\nloqcd$ + $\nloew$ + $\qcdqedps$ (blue lines), and LHE
(LHE$_{\alpha+\alpha_{\rm S}}$ in the plots, red lines) accuracy. With
``LHE'' we denote the partonic results as predicted by {\sc POWHEG}, after
the generation of the hard radiation, but before a complete PS is performed.
All the results in Fig.~\ref{fig:nlopslhe} include both QCD and EW
corrections. Results for $ZZ$ production are shown in the upper row, where we
consider the transverse momentum of the reconstructed $Z$ boson whose mass is
closer to $M_Z$, and we indicate it with $\pt(\bar{Z})$ (left plot), and the
invariant mass of the dimuon pair (right plot). In the lower row, we show
results for $WW$ production, where we consider the transverse momentum of the
positron (left plot) and the transverse momentum of the $e^+\mu^-$ pair
(right plot).

We start our discussion from the dimuon pair invariant mass. We observe a
relatively flat ratio between the fixed-order and the LHE result, slightly
shifted below one, for values of $M(\mu^+\mu^-)$ above the resonance peak,
whereas below the peak the LHE prediction is larger than the NLO one. These
trends are qualitatively similar to the $\nloew$/LO ratio $\delta_{\rm EW}$
shown in Fig.~\ref{fig:nloew}, and are mainly due to the radiative return,
although here they are numerically much smaller, since the difference between
the NLO and the LHE result comes essentially from the QED Sudakov form
factor. Adding the PS does not change the picture, as the PS radiation starts
at order $\alpha^2$ in the $\nloqcd$ + $\nloew$ + $\qcdqedps$ calculation.

The $\pt(e^+)$ and $\pt(\bar{Z})$ distributions are characterized by a
similar trend in the hard region, where the ratio between the LHE and the NLO
result is around 1.05 and is rather flat. The ratio of the showered and the
NLO result is qualitatively similar, and is about 1.1. The difference between
LHE and NLO+PS predictions is mainly due to QCD parton-shower radiation
recoiling against the positron (or the $\bar{Z}$ system).  At small
transverse momenta, the ratios under consideration exhibit different patterns
for the two observables. For the $\pt(\bar{Z})$ distribution, the LHE/NLO
ratio is essentially equal to one, whereas the NLO+PS result receives a
further (mild) suppression, due to multiple radiation. For the transverse
momentum of the positron, $\pt(e^+)$, the two ratio plots are qualitatively
similar to the one shown in Fig.~\ref{fig:nloqcd}. Since QED radiative
corrections play a smaller role (see Fig.~\ref{fig:nloew}), the shapes in
Fig.~\ref{fig:nlopslhe} are mainly due to QCD corrections. The size of the
effects displayed in Fig.~\ref{fig:nlopslhe} is significantly smaller than
the one at fixed order, for reasons analogous to those discussed for the
dimuon invariant mass, i.e.~the ratio plots in Fig.~\ref{fig:nlopslhe} expose
only higher-order corrections beyond the dominant ones.

Finally, we show the $\pt(e^+\mu^-)$ distribution, that allows us to further
elaborate on the origin of large QCD corrections for this process, and their
impact on a matched NLO+PS simulation thereof.  Below $\pt(e^+\mu^-)\simeq$
80 GeV, the LHE and the NLO+PS results are similar to each other, and they do
not show very large differences from the fixed-order result, except for a
moderate suppression when $\pt(e^+\mu^-)\to 0$. Instead, at larger
$\pt(e^+\mu^-)$ values, the LHE result has a tail harder than at NLO, and
this effect is even more marked at NLO+PS level. Similarly to what we have
already alluded to when discussing Fig.~\ref{fig:nlopsww}, distortion effects
for this distribution are due, by far, to the QCD corrections to diboson
production, and they can be understood as follows: although the
$\pt(e^+\mu^-)$ distribution is inclusive with respect to the QCD emissions,
it is kinematically correlated with the transverse momentum of the
color-singlet system (denoted $\pt(WW)$ in the following). In particular,
when $\pt(e^+\mu^-)$ approaches small values, the kinematics is dominated by
configurations where the color singlet is produced at small transverse
momentum. Here all-order effects from soft ISR dominate, yielding a Sudakov
suppression at small values of $\pt(WW)$, whose leftover is visible also in
$\pt(e^+\mu^-)$. For $\pt(e^+\mu^-)$ larger than 80--100 GeV, the dominant
kinematic configurations are those characterized by hard real QCD emissions.
We have already discussed that the main effect comes from a vector boson
recoiling against a hard jet, with the remaining vector boson being
relatively soft. A jet-veto can substantially reduce this enhancement,
keeping also the total $\nloqcd$/LO $K$-factor closer to one. However, in
Fig.~\ref{fig:nlopslhe} the ratios are taken with respect to the NLO result:
at relatively large values, the $\pt$ distribution is dominated by QCD real
radiation, and the ratio LHE/NLO shows a 20\% enhancement in this
region. This is a known {\sc POWHEG} effect for processes with relatively
large $K$-factors, first noticed in the context of inclusive Higgs boson
production in gluon fusion~\cite{Alioli:2008tz}.  It can be easily understood
by considering the expression of the {\sc POWHEG} differential cross section
of Eq.~(\ref{eq:dsigma}) at large transverse momentum. In this kinematic
region, the Sudakov form factor approaches 1, and the differential cross
section behaves like
\begin{eqnarray}
  d\sigma \approx \frac{\bar{B}(\Phi_B)}{ B (\Phi_B) } \left[  R_{\rm \sss QCD }(\Phi_B,
    \Phi_{\rm \sss rad})
    + R_{\rm \sss EW }(\Phi_B, \Phi_{\rm \sss rad})\right]  \, \mathd\Phi_B \,
  \mathd \Phi_{\rm \sss rad}\,,
  \nonumber
\end{eqnarray}
displaying an enhancement of a factor $\bar{B}/B$ with respect to the real
contribution.
We stress that this effect is formally subleading, although, for the case at
hand, amounts to about 20\%.  To reduce it, the use of {\tt hdamp} was
introduced in {\sc POWHEG}.
Essentially the real contribution is expressed as sum of two terms: one that
approaches the full real contribution in the soft and collinear limit, and
the other one that is finite in this limit. In
Eqs.~(\ref{eq:dsigma})--(\ref{eq:Sudakov}), only the singular term is used in
the generation of the hardest radiation by {\sc POWHEG}, while the other
term, being finite and positive, is included in the remnant contribution.
The scale {\tt hdamp} is used to separate the two terms.  In the present
simulation, no such separation of the real contribution has been performed,
so that in Eqs.~(\ref{eq:dsigma})--(\ref{eq:Sudakov}) the full real matrix
elements are used.

\begin{figure*}
  \begin{center}
    \includegraphics[scale=0.8]{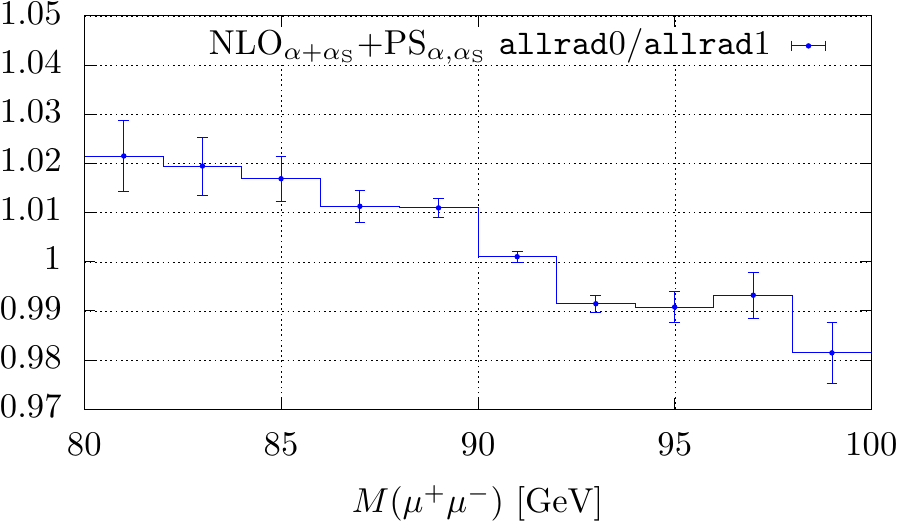}
    \includegraphics[scale=0.8]{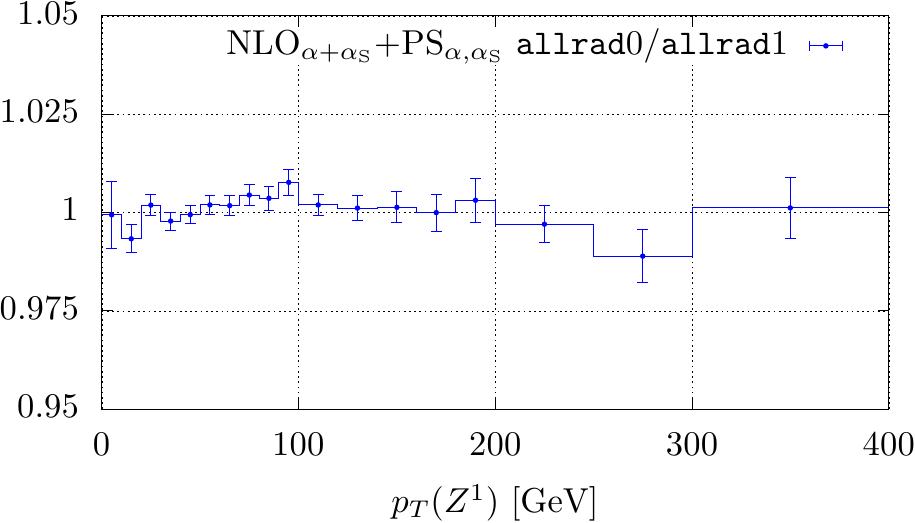}
    \caption{\label{fig:allrad} Ratio of the predictions at $\nloqcd$ +
      $\nloew$ + $\qcdqedps$ accuracy computed with and without the {\tt
        allrad} option ({\tt allrad}1 and {\tt allrad}0, respectively) for
      the invariant mass of the dimuon pair and the transverse momentum of
      the hardest $Z$ in $pp \to e^+e^-\mu^+\mu^-$. See main text for
      details.}
  \end{center}
\end{figure*}

The impact of the {\tt allrad} option for the invariant mass of
the dimuon pair and for the transverse momentum of the hardest $Z$ in $pp \to
e^+e^-\mu^+\mu^-$ is displayed in Fig.~\ref{fig:allrad}. We find an effect of
about 1-2\% below the $Z$ resonance for the $M(\mu^+\mu^-)$ distribution,
where the QCD corrections are a flat normalization factor, while the QED
corrections lead to large shape distortions.
This is due to the fact that, without the {\tt allrad} option, the upper
scale for the generation of radiation used by the PS program is set by the
transverse momentum of the hardest radiation generated by {\sc POWHEG}, that,
in most cases, is ISR QCD. This upper scale is then used, by the PS, for both the ISR
(QCD and QED) and the FSR QED radiation.
As a result, the FSR QED radiation from PS can be harder than the one that
would have been generated by {\sc POWHEG} and the predictions obtained
without the {\tt allrad} option show larger radiative return effects compared
to the ones computed with the {\tt allrad} option. At variance with the
invariant-mass distribution, for the $\pt$ of the hardest Z, where the QCD
corrections are much larger than the EW ones, the effect of the {\tt allrad}
option is not visible with the statistics used in this paper. The results
shown in Fig.~\ref{fig:allrad} agree with the ones obtained in
Ref.~\cite{CarloniCalame:2016ouw} for Drell-Yan production.

Before drawing our conclusions, we would like to comment further on the
possible sources of theoretical uncertainties that affect our calculation. We
stress that, as far as the perturbative uncertainties are concerned, all
these effects are of higher orders, beyond the $\nloqcd$ + $\nloew$ + PS
accuracy of the presented results.  The main source of theoretical
uncertainties are the perturbative ones related to the choice of the
renormalization and factorization scales, and the non-perturbative ones
coming from the choice of the parton distribution functions. It is possible
to estimate these uncertainties through scale variations and using several
PDF sets. The impact of scale variation at $\nloqcd$ for the processes under
consideration can be found, for instance, in Ref.~\cite{Chiesa:2018lcs}.  The
theoretical uncertainties associated with the missing higher-order EW
corrections are expected to have a minor impact and can be estimated by
changing the EW input parameters/renormalization schemes using the flag {\tt
  scheme} in the input card to select the $(\alpha_0,M_W,M_Z)$,
$(\alpha(M_Z),M_W,M_Z)$, or $(G_\mu,M_W,M_Z)$ schemes.

Photon-induced processes contribute at the same perturbative order of the
$\nloew$ corrections (for $WW$ and $ZZ$ they also contribute at LO): at
present, these processes are not included in our calculation and their
contribution is a theoretical uncertainty for the results presented in this
paper.
The study of their effect in a NLO+PS simulation will be the subject of
future investigations 

As stated in Sect.~\ref{sec:intro}, the {\sc POWHEG} algorithm applied to the
case of $\nloqcd$+$\nloew$ corrections gives as a by-product mixed
corrections of the form $\mathcal{O}(\as^n \alpha^m)$.  This is an
approximation of the factorizable mixed QCD-EW corrections that is only
reliable in the collinear limit. For diboson production processes, this
approximation is expected to work for distributions like the vector-bosons
virtualities in the regions close to the resonances (similarly to what
happens for Drell-Yan~\cite{CarloniCalame:2016ouw}), while it is expected to
fail in the high-$\pt$ regions, where the QCD corrections are very large and
dominated by hard QCD radiation~\cite{Kallweit:2019zez}.  Although the
reliability of the mixed effect provided by {\sc POWHEG} can only be assessed
through comparison with full calculations at $\mathcal{O}(\alpha\alpha_S)$,
when such calculations will be available, an estimate of the related
uncertainties can be obtained by comparing the results at $\nloqcd$ +
$\nloew$ + PS accuracy for diboson and diboson\,+\,jet production (possibly
merged with the results for $pp\to VV'$).

Another class of uncertainties is related to the matching procedure, such as
the scale choice for the parameter {\tt hdamp}, or the use of the {\tt
  allrad} option.  In addition, different PS algorithms use different
ordering variables, and such choices have a non negligible impact on the
kinematic distributions.  The present release of the code presented in this
paper only includes an interface to {\sc PYTHIA8.2}, however, in order to
study this class of uncertainties, dedicated interfaces to other shower Monte
Carlo programs (like, for instance, {\tt PHOTOS}~\cite{Barberio:1990ms,
  Barberio:1993qi, Golonka:2005pn} for the QED FSR) could be developed.

A comprehensive study of the theoretical uncertainties described above goes
beyond the scope of this work, however, the event generator presented in this
paper could be used in particular by the experimental collaborations to asses
the different classes of theoretical uncertainties that affect the
calculation at hand.

\section{Conclusions}
\label{sec:conclusions}

We computed the NLO QCD + NLO EW corrections to diboson production at hadron
colliders matched to a complete parton shower, where QCD and QED radiation is
simulated. For diboson production this is the first calculation where the NLO
EW corrections have been consistently matched to QED PS.
As the considered processes involve the production and the decay of unstable
particles, whose decay products can radiate photons, the calculation is based
on the \RES{} framework.  The corresponding code is public and all the
information for downloading it can be found in the \BOX{} web
page.\footnote{\url{http://powhegbox.mib.infn.it/}}

Though we did not perform an extensive phenomenological study that might be
the subject of a future publication, we showed the potential of our code, and
we pointed out that EW effects, consistently matched to QED parton shower,
are relevant for several observables of interest.

The code relies on the {\sc Recola2} library for the calculation of the
matrix elements for four leptons and four leptons plus photon/parton
production. In particular, we developed a fully general interface between
\BOX{} and {\sc Recola2} that could be used for other processes. We performed
our calculation in the Standard Model.  However, given the possibility of
{\sc Recola2} to compute tree-level and one-loop matrix elements in general
extensions of the SM, in the future the code could be easily generalized to
compute the NLO QCD + NLO EW corrections to diboson production matched to QCD
and QED parton shower in the context of models beyond the SM, provided that
the one-loop corrections are available in {\sc Recola2}, and that the
considered model does not alter the structure of QED and QCD interactions in
a non-trivial way. If this were the case, the subtraction of infrared
singularities and the Sudakov form factors in the \RES{} should also be
generalized.

As a final remark, the effect of NLO QCD corrections to diboson production is
very large and it is dominated by real parton radiation, especially in the
regions where one of the two vector bosons is soft with respect to the
jet. It is thus important to include QCD corrections beyond NLO accuracy, for
instance using a consistent merging of the predictions for $VV'$ and $VV'$+
jet production ($V,V'=W,Z$) based on the {\sc MiNLO}~\cite{Hamilton:2012rf}
or {\sc MiNNLO$_{\rm PS}$}~\cite{Monni:2019whf} procedures (at NLO and NNLO
accuracy, respectively). The code presented in this paper can be taken as the
starting point for the inclusion of the EW effects matched to QED PS in the
treatment of diboson (+\,jet) production in the {\sc MiNLO} or {\sc
  MiNNLO$_{\rm PS}$} framework.

\subsection*{Acknowledgements}
The work of M.C.~has been supported by the ``Investissements d’avenir, Labex ENIGMASS''.

\bibliographystyle{spphys}       
\bibliography{diboson}   

\end{document}